\documentclass[
 reprint,
 superscriptaddress,
 groupedaddress,
showpacs,
preprintnumbers,
citeautoscript,
amsmath,
amssymb,
aps,
prb,
]{revtex4-1}

\usepackage{soul}
\usepackage{xcolor}
\usepackage{graphicx}
\usepackage{dcolumn}
\usepackage{bm}
\usepackage{color}
\usepackage{comment}
\usepackage{hyperref}
\usepackage[mathlines]{lineno}
\usepackage{mathtools}
\usepackage{amsmath}
\usepackage{amssymb}
\graphicspath{{figures/}} 

\usepackage{braket}
\usepackage{multirow}
\usepackage{array}

\begin{document}

\title{Non-equilibrium BN-ZnO: Optical properties and excitonic effects from first principles}  

\author{Xiao Zhang}
\affiliation{Department of Mechanical Science and Engineering, University of Illinois at Urbana-Champaign, Urbana, IL 61801, USA}

\author{Andr\'e Schleife}
\email{schleife@illinois.edu}
\affiliation{Department of Materials Science and Engineering, University of Illinois at Urbana-Champaign, Urbana, IL 61801, USA}
\affiliation{Frederick Seitz Materials Research Laboratory, University of Illinois at Urbana-Champaign, Urbana, IL 61801, USA}
\affiliation{National Center for Supercomputing Applications, University of Illinois at Urbana-Champaign, Urbana, IL 61801, USA}

\date{\today}

\begin{abstract}
The non-equilibrium boron nitride (BN) phase of zinc oxide (ZnO) has been reported for thin films and nanostructures, however, its properties are not well understood due to a persistent controversy that prevents reconciling experimental and first-principles results for its atomic coordinates.
We use first-principles theoretical spectroscopy to accurately compute electronic and optical properties, including single-quasiparticle and excitonic effects:
Band structures and densities of states are computed using density functional theory, hybrid functionals, and the $GW$ approximation.
Accurate optical absorption spectra and exciton binding energies are computed by solving the Bethe-Salpeter equation for the optical polarization function.
Using this data we show that the band-gap difference between BN-ZnO and wurtzite (WZ) ZnO agrees very well with experiment when the theoretical lattice geometry is used, but significantly disagrees for the experimental atomic coordinates.
We also show that the optical anisotropy of BN-ZnO differs significantly from that of WZ-ZnO, allowing to optically distinguish both polymorphs.
By using the transfer-matrix method to solve Maxwell's equations for thin films composed of both polymorphs, we illustrate that this opens up a promising route for tuning optical properties.
\end{abstract}

\maketitle

\section{\label{sec:intro}Introduction}

Zinc oxide (ZnO) is a semiconductor that keeps attracting strong interest both from a fundamental and applied point of view:
It is an earth-abundant, non-toxic material with a large band gap, rendering it transparent to visible light.
ZnO can be heavily $n$-doped to achieve high free-carrier concentrations, leading to conductivities that are suitable for semiconductor devices \cite{ueda2001}.
It has been widely used as $n$-type conductor in heterostructures, e.g.\ in light-emitting and photo diodes \cite{ozgur05} and, potentially, as 
transparent transistor \cite{fort12}.
More recently, nanostructures of ZnO, $n$-doped with Al, are  becoming increasingly important for devices since they can be used in high-performance transparent electrodes, e.g.\ for solar cells \cite{sun2006synthesis}.

\begin{figure}
\centering
\includegraphics[width=0.45\textwidth]{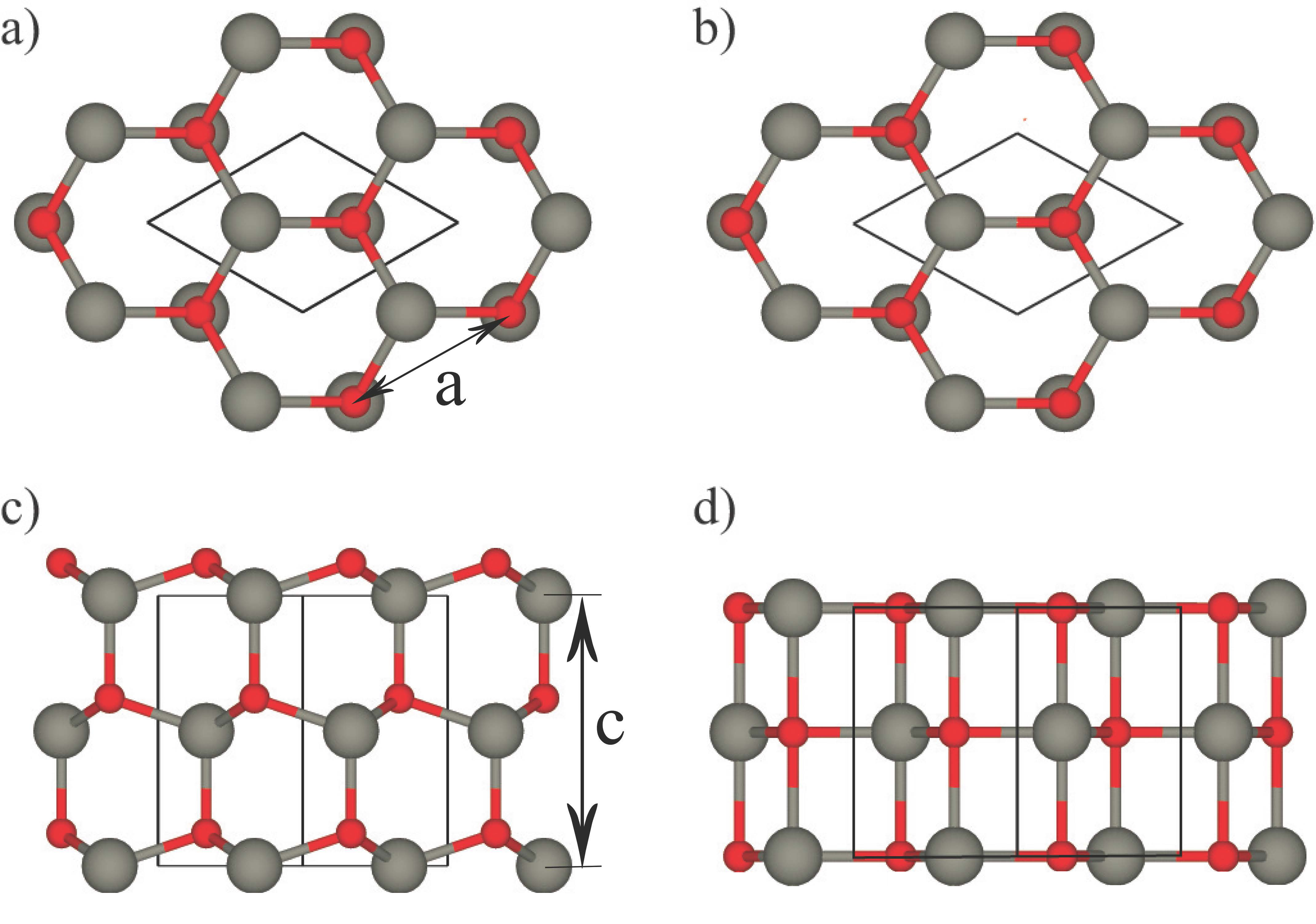}
\caption{\label{fig:strs}(Color online.) Structure of WZ-ZnO (left) and BN-ZnO (right). (a) and (b) show a view from the $[000\bar{1}]$ and (c) and (d) a view from the $[11\bar{2}0]$ direction. Red spheres represent oxygen atoms and gray spheres zinc atoms. From (a) and (b), one can see that the two polymorphs have the same hexagonal crystal structure. From (c) and (d), one can see that the BN phase has planar stacking in the $[0001]$ direction instead of a Zn/O centered tetrahedral configuration.
}
\end{figure}

However, despite intensive research on this material, important properties, in particular of its nanostructures, remain poorly understood.
Most prominent amongst those is the phase stability under non-equilibrium preparation conditions:
In equilibrium under ambient conditions, bulk ZnO crystallizes in the wurtzite (WZ) structure  \cite{zhu08}.
High-pressure zinc-blende or rock-salt phases were predicted by theory \cite{zag14} and observed in experiment \cite{recio1998compressibility, segura2003optical, chichvarina2015stable}. 
One of these non-equilibrium phases, the boron-nitride (BN) structure (P6$_3$/mmc; space group no.\ 194), stands out:
Similarly to WZ it is based on a hexagonal lattice, but the basis atoms along the $c$ direction are arranged to form stoichiometric planes (see Fig.\ \ref{fig:strs}). 

While in bulk ZnO the BN polymorph is only stable for large compressive or tensile deformation, this changes under non-equilibrium conditions used during synthesis of low-dimensional geometries such as nanorods or thin films:
BN-ZnO was reported to appear in nano-structured ZnO by both experiment \cite{tusche07,claeyssens2005growth, wu2011stabilizing} and theory \cite{claeyssens2005growth,freeman2006graphitic,kulkarni06novel}.
In particular, ultra-thin films composed of 2\,--\,5 mono-layers, e.g.\ on metal substrates such as Ag or Pd, grow as BN-ZnO \cite{tusche07,weirum2010growth, liu2015stabilization}.
It is also predicted that the BN structure appears in ZnO nanowires with lateral dimensions of around 4 nm\cite{kulkarni06novel} under tensile strain along the wire direction and that the BN polymorph can be stabilized in thicker films (up to 48 mono-layers) if grown with lattice mismatch to provide tensile strain \cite{wu2011stabilizing}.

Furthermore, it is reported that the BN phase affects the mechanism of ZnO thin-film growth:
The polar (0001) surface energy is higher than that of the non-polar $(10\bar{1}0)$ and $(11\bar{2}0)$ surfaces for WZ-ZnO, rendering growth of this surface energetically unfavorable.
However, in thin films up to 9 mono-layers, the (0001) BN-ZnO surface is energetically more stable than these three WZ surfaces and it is relatively easy to achieve a transition from BN-ZnO to WZ-ZnO as thicker films grow.
Hence, even though the first few layers of ZnO grown on a substrate may appear in the BN phase, the WZ phase will still dominate the morphology of thicker samples.
These will then show the (0001) geometry, even though that surface energy is higher than those of $(10\bar{1}0)$ and $(11\bar{2}0)$ surfaces of WZ-ZnO \cite{claeyssens2005growth}.
In addition, in thin films the BN structure is a possible route to circumvent strong dipoles that were reported for (0001) surfaces of WZ-ZnO \cite{claeyssens2005growth,freeman2006graphitic,liu2015stabilization}.
Since strong dipoles make the surface electronically unstable \cite{mora2017polar,staemmler2003stabilization}, circumventing the surface dipole is important for growing (0001) WZ-ZnO surfaces.

WZ-ZnO and BN-ZnO contributions to thin films and nanostructures are important for optoelectronic applications, since this opens an interesting possibility for tuning optical properties by phase selectivity.
Unfortunately, the optical properties of BN-ZnO have never been explained in detail.
We attribute this to a persistent controversy that, although receiving considerable theoretical and experimental attention, has never been resolved:
Pueyo \emph{et al.}\ \cite{liz10} reported a wet-chemical route to grow bulk BN-ZnO under specific chemical conditions. 
Subsequent structural analysis showed that first-principles simulations significantly overestimate lattice parameters \cite{liz10, bip11, zag14}.
At the same time, the band gap of BN-ZnO determined in experiment is only slightly larger than that of WZ-ZnO.
This, again, contradicts first-principles results that find such a small gap difference \emph{only} when theoretical lattice parameters are used \cite{liz10}.
Electronic and optical properties of BN-ZnO have never been reconciled with the underlying lattice geometry.

Fortunately, highly accurate theoretical-spectroscopy techniques can now directly provide experiment with guidance needed to reliably distinguish different phases \emph{optically}, even in nanostructures \cite{Lim:2017}.
Here we combine first-principles simulations with Maxwell modeling to derive an optical signature that can be used to unambiguously identify BN contributions in ZnO samples using optical-absorption experiments.
To this end, we study structural, electronic, and optical properties of bulk BN-ZnO and WZ-ZnO using density functional and many-body perturbation theory.
We show that the theoretical lattice constants lead to a consistent picture of electronic structure and optical properties, in agreement with experiment.
In particular, BN-ZnO shows a larger band gap and, due to optical anisotropy, a two-step onset of optical absorption.
By using these results as input to Maxwell simulations of thin ZnO films, we show that the emerging line-shape difference of the optical-absorption onset allows to clearly and unambiguously distinguish between BN-ZnO and WZ-ZnO in optical measurements.
The significantly increased transmission in the UV spectral region that is important for applications of ZnO nanostructures \cite{alenezi2015chip,alchaar2017enhanced,chi2017ultra}, can explain experimental observations, that were not compatible with optical absorption of WZ-ZnO.
In addition, our results help designing optical properties of ZnO films by purposefully combining both phases.

The remainder of this work is structured as follows:
In Sec.\ \ref{sec:method} the theoretical framework and computational approach is described.
Lattice geometries, total energies, and transition pressures are discussed in Sec.\ \ref{sec:struc}.
Results for band structures are reported in Sec.\ \ref{sec:band} and for optical properties and exciton-binding energies in Sec.\ \ref{sec:opt}.
Finally, Sec.\ \ref{sec:conclusion} concludes this work.

\section{\label{sec:method}Theoretical and computational approach}

Density functional theory (DFT) \cite{hohen64} is a reliable, widely used technique to study ground-state properties of materials from first principles \cite{argaman00}.
Within the Kohn-Sham (KS) approach \cite{ks65} the fully interacting electronic system is mapped to a non-interacting system in an effective potential.
In this work we use a generalized-gradient approximation (GGA) \cite{per92} to describe exchange and correlation (XC).
The electron-ion interaction is described by means of the projector-augmented wave (PAW) scheme \cite{blo94}, which allows using a small cutoff energy for the plane-wave expansion of the wave functions, while still achieving accurate self-consistent results.
After testing for convergence, we use 450 eV for the cutoff energy and employ a $12\,\times\,12\,\times\,8$ $\Gamma$-centered Monkhorst-Pack (MP) $\mathbf{k}$-point grid \cite{Monkhorst_1976} to sample the Brillouin zone for both the BN and WZ structure.
Relaxation of atomic coordinates according to Hellman-Feynman forces provides us with the equilibrium lattice structure for both polymorphs.

The electronic structure cannot be described using KS eigenvalues from DFT-GGA, since those significantly underestimate the band gap of semiconductors \cite{sham1983}.
In order to compute accurate band gaps we rely on many-body perturbation theory and the quasiparticle approximation \cite{onida2002electronic}.
This technique achieves a reliable description of single-particle excitations by taking the effect of adding or removing an electron from the system into account.
Here we specifically compute quasiparticle energies within Hedin's $GW$ approximation \cite{hedi65}, that provides excited-state properties, such as band structures and densities of states, that are directly comparable to experiment. 
In the $GW$ framework, the following Dyson equation is solved\cite{ccmp06},
\begin{equation}
G=G_0+G_0\Sigma G,
\end{equation}
where $G_0$ is the Green's function of the non-interacting system and $G$ is the full Green's function.
$\Sigma$ is the self-energy term that contains all many-body XC effects.
In $GW$ approximation, this term is approximated to be $\Sigma$=$iGW$, where $W$ is the screened Coulomb interaction of electrons.

Usually, the quasiparticle Dyson equation is not solved fully self-consistently.
Due to the high computational cost, in practice a perturbation-theory approach is used and corrections with respect to the starting electronic structure are computed.
Hence, if only one or few steps of perturbation theory are used and the starting point is too far from the final result, a remaining dependence of the electronic structure on the starting point is observed.
In our work, this effect is considered by comparing quasiparticle energies from one-step perturbation theory ($G_0W_0$), using $G$ and $W$ computed from  KS wave functions, to quasiparticle energies computed after updating both $G$ and $W$ up to nine times (sc$GW$).
The difference of the band gap between the last two iterations is then smaller than 0.01 eV.

In order to mitigate the starting-point dependence at lower computational cost, we also compare to using a hybrid XC functional to compute the input electronic structure for $G_0W_0$ calculations.
In the hybrid functional approach, a certain fraction of Hartree-Fock exact exchange is combined with the GGA XC functional.
We specifically use the HSE06 functional by Heyd, Scuseria, and Ernzerhof \cite{hse,heyd2006erratum} which often gives a much improved description of the electronic structure.
The large computational cost of HSE06 and $GW$ calculations requires us to use a smaller $8\,\times\,8\,\times\,6$ MP $\mathbf{k}$-point grid.

Finally, we compute optical absorption spectra of WZ- and BN-ZnO and take excitonic effects into account.
An exciton is a quasiparticle used to describe Coulomb-bound electron-hole pairs in optically excited states \cite{toyozawa68}.
Excitonic effects reduce the minimum excitation energy of valence-conduction band excitations compared to non-interacting electron-hole pairs.
The energy difference between \emph{optical} gap and single-quasiparticle gap is the exciton binding energy, which is reported to be about 65 meV for WZ-ZnO \cite{look01}.

In order to take excitonic effects into account, we solve the Bethe-Salpeter Equation (BSE) \cite{salpeter1951} for the optical polarization function.
In practice, the BSE is transformed to an eigenvalue problem\cite{marini2009yambo} for the exciton Hamiltonian,
\begin{equation}
\begin{split}
H_{cv\mathbf{k},c'v'\mathbf{k'}}=&(\epsilon_{c\mathbf{k}}-\epsilon_{v\mathbf{k'}})\delta_{cc'}\delta_{vv'}\delta_{\mathbf{kk'}}\\
&+2v^{v'c'\mathbf{k}'}_{vc\mathbf{k}}-W^{v'c'\mathbf{k}'}_{vc\mathbf{k}},
\end{split}
\end{equation}
where the diagonal contains single-quasiparticle excitation energies.
Subscripts $c$, $v$, and $k$ denote indices of conduction bands, valence bands, and $\mathbf{k}$-points. 
The screened ($W$) and unscreened ($v$) Coulomb potentials describe the electron-hole attraction and local-field effects, respectively.
Coulomb matrix elements are evaluated using KS states from DFT-GGA.
Quasiparticle energies in the exciton Hamiltonian are approximated using a rigid scissor shift of 2.26 eV for BN-ZnO and 2.32 eV for WZ-ZnO to open up the band gap of DFT-GGA to the HSE06+$G_0W_0$ values.
After testing for convergence of optical spectra, we increase the $\mathbf{k}$-point sampling to $20\,\times\,20\,\times\,14$ and include a small random shift.
This provides well-converged optical spectra for the BN and WZ polymorphs.

All calculations in this work are carried out with the Vienna \emph{Ab-Initio} Simulation Package \cite{Gajdos:2006,Kresse:1999,kresse96,shishkin2007self, shishkin2007accurate} (VASP) and the BSE implementation discussed in Refs.\ \onlinecite{Roedl:2008,fuchs2008efficient}.
All relevant input and output files and source data for generating figures and tables in this paper can be found in the Materials Data Facility \cite{MDF,data}. 

\section{\label{sec:struc}Structural properties and transition pressure}

\subsection{\label{sec:relax}Atomic geometries}

\begin{table}
\begin{center}
\caption{\label{tab:fitresult}Parameters of the equation of state fit.}
\begin{tabular}{ c c c c c }
\hline
structure   & $V_0$(\r{A}) & $E_0$ (eV)   & $B_0$ (GPa) & $B'_0$  \\ \hline
WZ & 49.41   & $-18.09$   &  132  & 4.68     \\ 
BN & 47.54   & $-17.81$  & 122   & 7.46       \\ \hline
\end{tabular}
\end{center}
\end{table}

\begin{table}
\caption{\label{tab:str}Lattice parameters of WZ and BN ZnO.}
\begin{tabular}{cccc}
\hline
WZ & $a$ (\AA) & $c$ (\AA) &$c/a$ \\ \hline
This work (GGA) & 3.281 & 5.300 & 1.615\\
LDA\footnote{\label{zag14}Ref.\ \onlinecite{zag14}} & 3.190 & 5.180 &1.624\\
LDA\footnote{\label{jaffe}Ref.\ \onlinecite{jaffe2000lda}} &3.199 & 5.162&1.614\\
GGA\footnotemark[2] & 3.292 &5.292 &1.608\\
B3LYP\footnotemark[1] & 3.280 & 5.290 &1.613\\
Hartree-Fock\footnotemark[1] & 3.290 & 5.240 &1.593\\ 
Expt.\footnote{Ref.\ \onlinecite{kar96}}&3.245&5.204&1.604\\ \hline
\end{tabular}
\begin{tabular}{cccc}
\hline
BN & $a$ (\AA) & $c$ (\AA) & $c/a$ \\ \hline
This work (GGA)  &  3.455 & 4.599 & 1.331\\
LDA\footnote{Ref.\ \onlinecite{bip11}} & 3.370 & 4.460 & 1.323\\
LDA\footnotemark[1] &3.400&4.390 &1.291\\
GGA\footnotemark[4] &3.450&4.620 & 1.339\\
LDA+$U$\footnotemark[4] & 3.190 &3.970 & 1.245\\
GGA+$U$\footnotemark[4]&3.300&4.180 & 1.267\\
B3LYP\footnotemark[1] & 3.480 & 4.540 &1.305\\
Hartree-Fock\footnotemark[1]&3.480 &4.460 &1.282\\ 
Expt.\footnote{Ref.\ \onlinecite{liz10}}& 3.099&3.858 &1.245\\ \hline
\end{tabular}
\end{table}

We computed fully relaxed atomic structures for the WZ and BN polymorphs of ZnO using DFT-GGA.
First, we compute total energies for several unit-cell sizes within 3\,\% of the equilibrium value and fit the Murnaghan equation of state \cite{murnaghan44}, in order to obtain the equilibrium volume, total energy, and bulk modulus.
These results are summarized in Table \ref{tab:fitresult} (see details in the supplemental material at [URL will be inserted by publisher]).
We then performed a final relaxation to make sure that the Hellmann-Feynman forces acting on each atom are smaller than 5 meV/\r{A}. 
The final results for the $a$ and $c$ lattice parameters are compiled in Table \ref{tab:str}, along with computational \cite{zag14, bip11,jaffe2000lda} and experimental data \cite{kar96, liz10} from the literature.

This data shows excellent agreement of our results with other computational reports.
For both polymorphs the DFT-GGA values reported by Rakshit \emph{et al.}\ \cite{bip11} agree within 1\,\%.
Data from calculations using a local-density approximation (LDA) shows smaller lattice constants for BN-ZnO and WZ-ZnO (see Table \ref{tab:str}) compared to DFT-GGA.
This is expected, since LDA usually overbinds by a few percent (i.e.\ underestimates lattice constants) and GGA usually underbinds, compared to experiment.
Computational results \cite{zag14} from hybrid functionals differ by less than 3\,\% from our GGA data for both polymorphs.

Comparison with experiment shows a more complicated picture and illustrates the controversy discussed above.
While DFT-GGA overestimates experimental lattice constants of WZ-ZnO by less than 2\,\% (see Table \ref{tab:str}), the agreement is \emph{significantly} worse for BN-ZnO:
In this case DFT-GGA overestimates $a$ and $c$, obtained using powder X-ray diffraction \cite{liz10} (see Table \ref{tab:str}), by 11.5\,\% and 19\,\%, respectively.
Not only is this an unusually large deviation, but more importantly, even DFT-LDA, which normally overbinds, still overestimates the experimental results by more than 13\,\%.

Rakshit \emph{et al.}, who observe a similar behavior, relate the incorrect atomic geometries to too shallow Zn\,3$d$ states when GGA or LDA are used \cite{bip11}.
They showed that agreement within 6\,--\,8\,\% can be achieved by means of a GGA+$U$ XC functional, which significantly lowers the Zn\,3$d$ states in energy.
They also applied LDA+$U$ using 12 eV for the $U$ value, which is much larger compared to what is typically reported for WZ-ZnO (7\,--\,9 eV, see Ref.\ \onlinecite{karazhanov2007electronic}).
This lowers the Zn\,3$d$ electrons in energy to about 8\,--\,10 eV below the valence-band maximum (VBM), which is lower than experimental results that report Zn\,3$d$ states for WZ-ZnO at about 8 eV below the VBM \cite{preston2008band}.
However, even in this case DFT-LDA still overestimates lattice parameters by about 3\,\%, instead of overbinding \cite{bip11}.
In addition, it remains unexplained why that same effect would not influence the WZ polymorph of ZnO, where the Zn\,$3d$ states are equally shallow in DFT-LDA or DFT-GGA.
Hence, at this point we conclude that the unexpectedly small lattice parameters of BN-ZnO reported from experiment \cite{liz10} are not fully understood.
In Secs.\ \ref{sec:band} and \ref{sec:opt} we combine theoretical and experimental data for the lattice geometries with theoretical spectroscopy to shed further light on this problem:
We show that the experimentally reported lattice geometry is inconsistent with reported electronic band gaps.

\subsection{\label{sec:trp}Transition pressures}

\begin{figure}
\begin{center}
\includegraphics[width=0.85\columnwidth]{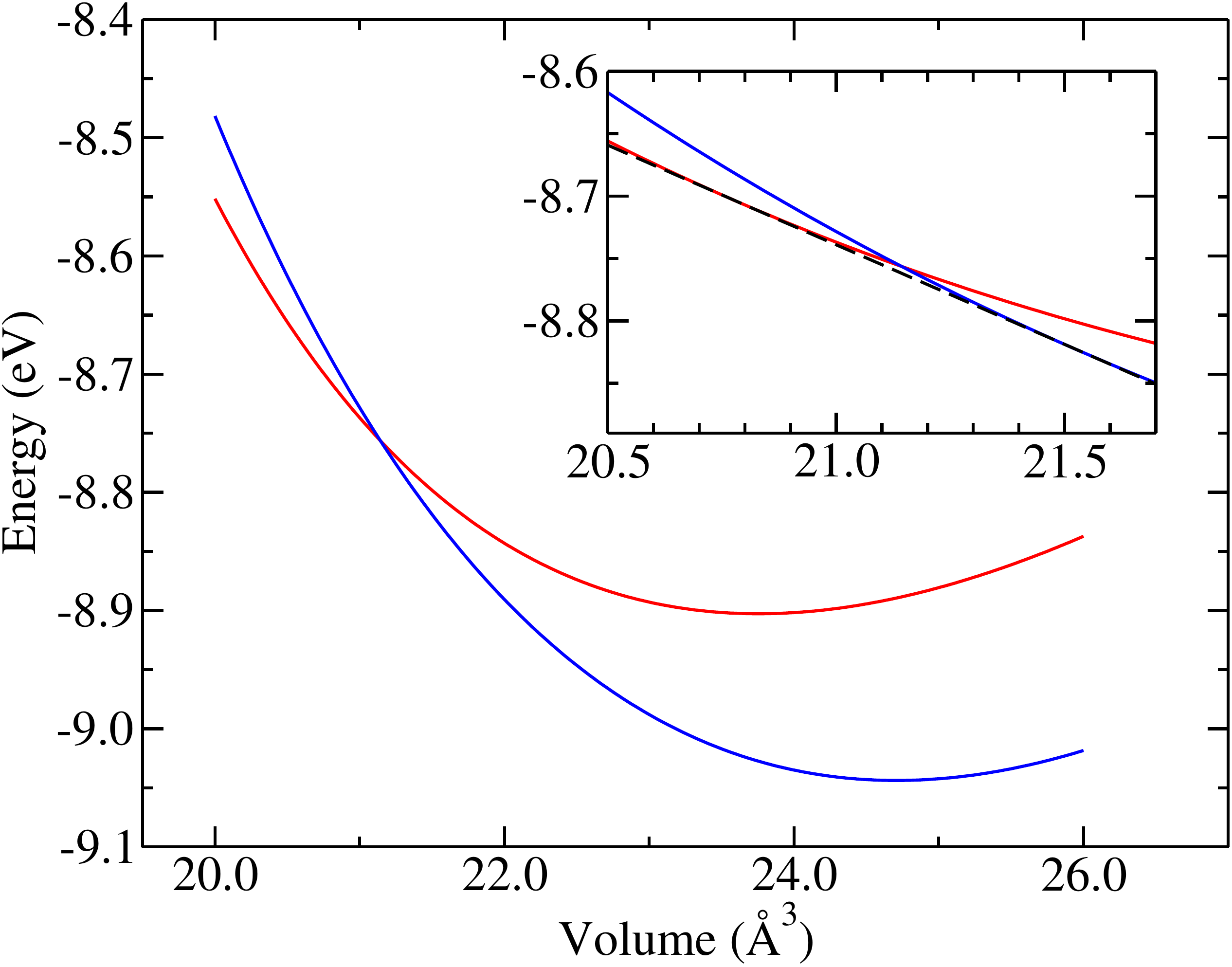}
\end{center}
\caption{\label{fig:trp}(Color online.) (a) Total energy of ZnO (in eV) as a function of the unit-cell volume (in \AA$^3$), both per formula unit. Data is shown for WZ-ZnO (blue) and BN-ZnO (red). The inset magnifies the intersection point and shows the common tangent line (dashed). 
}
\end{figure}

By applying the common-tangent method to the $E(V)$ curves of the two different structures (see Fig.\ \ref{fig:trp}), we compute the pressure for the phase transition from WZ- to BN-ZnO to be 25.6 GPa.
We confirm this result by computing the enthalpy, $H$=$E$+$pV$, for both phases (data not shown), where $V$ is the volume of the unit cell, $E$ is the corresponding total energy, and $p$ is the external pressure.
By changing the volume of the unit cell and, thus, the external pressure, we determine the transition pressure from the crossing point of the enthalpies of both polymorphs.
The thermodynamically stable polymorph minimizes the enthalpy at a given pressure.
The results from both methods agree with each other.

Zagorac \emph{et al.}\ used a hybrid XC functional and found another phase transition from WZ to BN-ZnO to occur under negative pressure. We did not study this regime further in this work.

\section{\label{sec:band}Electronic structure}

\begin{figure}
\includegraphics[width=0.48\textwidth]{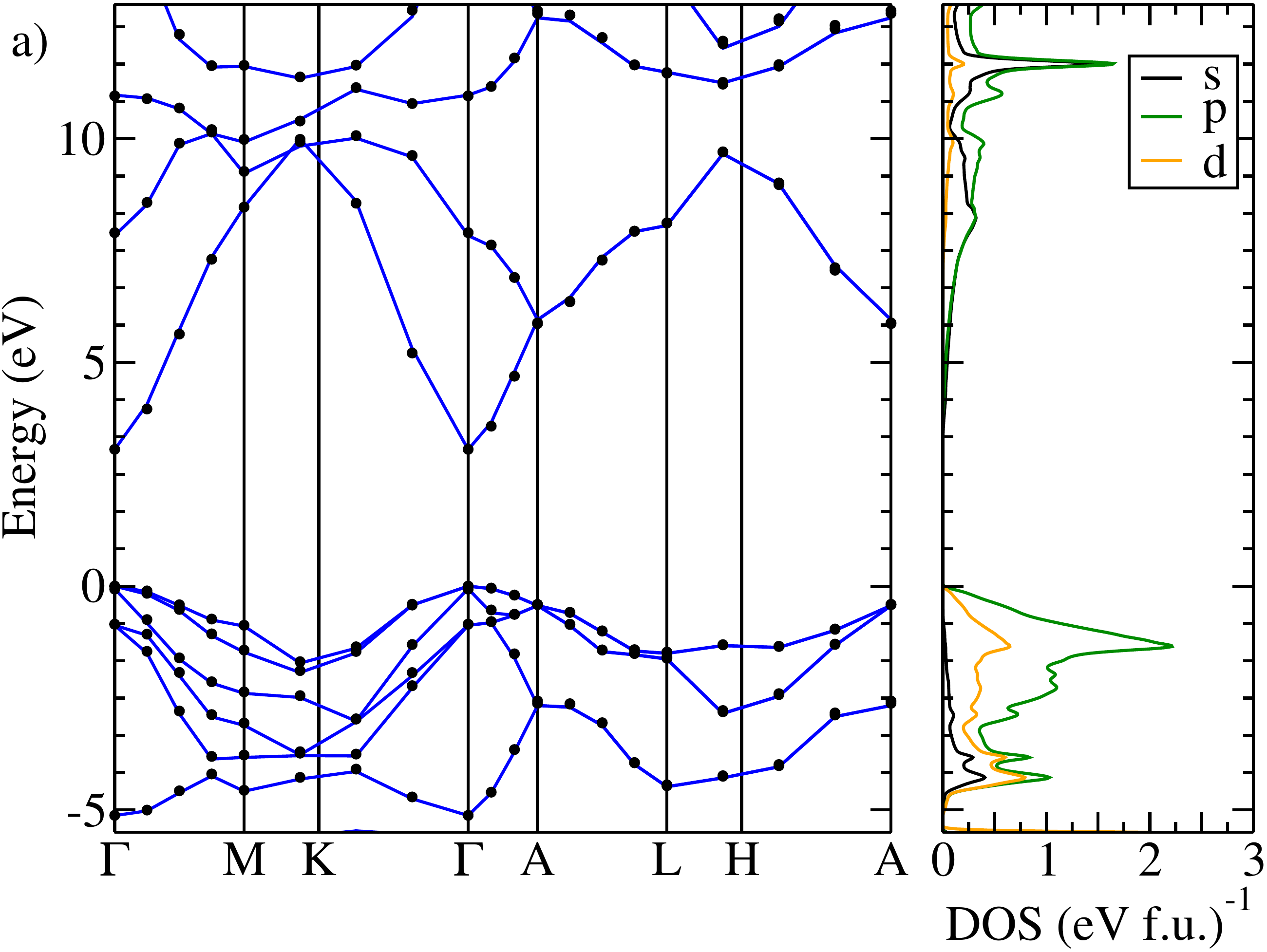}
\includegraphics[width=0.48\textwidth]{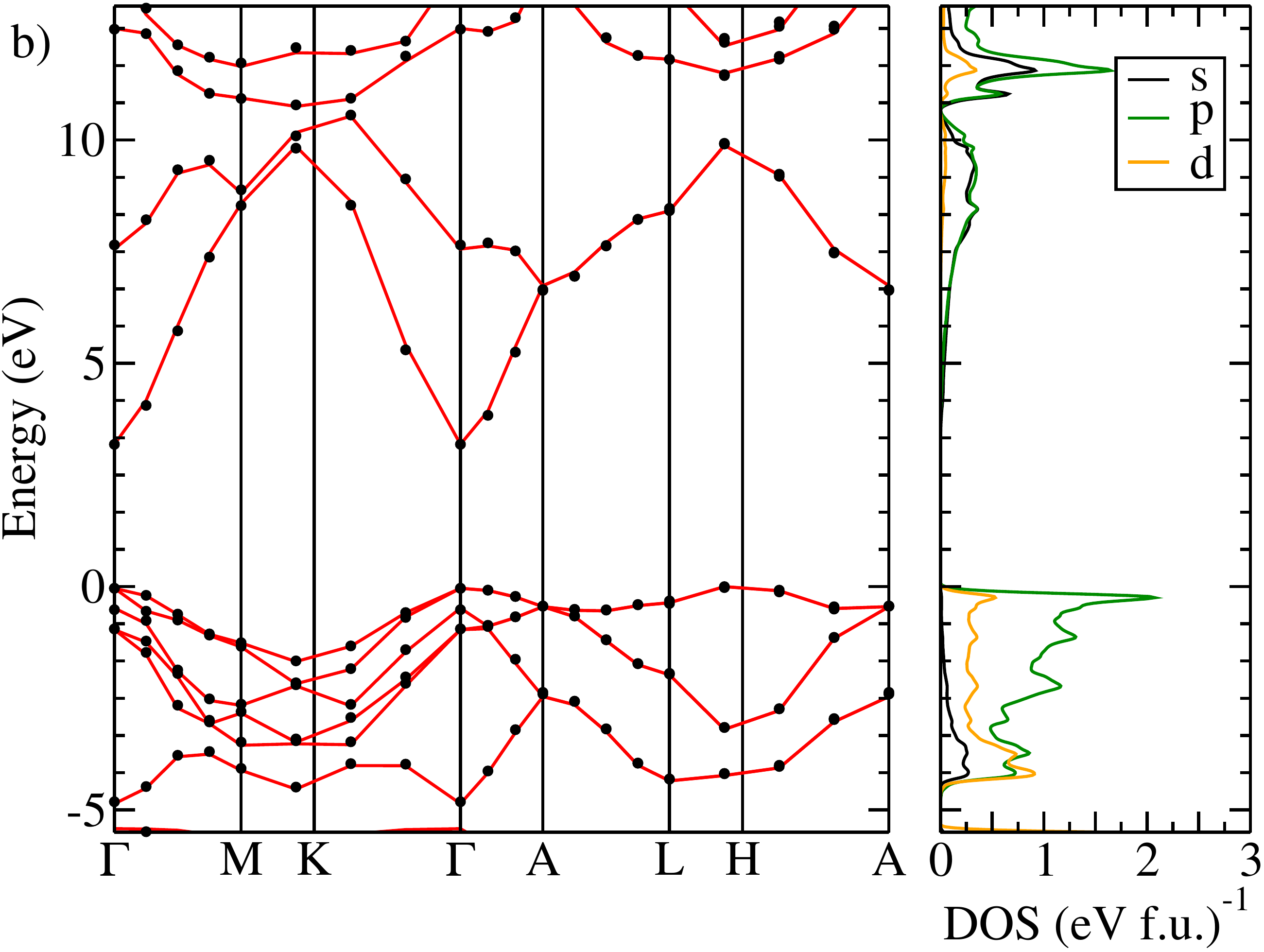}
\caption{\label{fig:GWband}(Color online.) The band structure and density of states for (a) WZ-ZnO and (b) BN-ZnO, computed using the HSE06+$G_0W_0$ approach (red and blue lines). Black dots represent the results from HSE06+sc$GW$, after adjusting the gap to the HSE06+$G_0W_0$ value.
}
\end{figure}

\begin{table}
\begin{center}
\caption{\label{tab:band}Band gaps (in eV), computed using different methods, compared to experiment. ``BN (theor.)'' and ``BN (Exp.)'' refer to the lattice constants used in the calculations.}
\begin{tabular}{cccc}
\hline
  & WZ & BN (theor.) & BN (Exp.)\\ \hline
This work (GGA)&0.74&0.90&1.93 \\
This work (HSE06) &2.40 &2.60 & 3.82\\
This work (GGA+$G_0W_0$) & 2.10 & 2.35 &3.89\\
This work (GGA+sc$GW$) & 3.85 & 3.96 &\\
This work (HSE06+$G_0W_0$) &3.05 &3.22 &4.61\\
This work (HSE06+sc$GW$) &3.97 & 4.04 & 5.36 \\
GGA\footnote{Ref. \onlinecite{wrobel2009calculations}}& 0.80 &&\\
GGA\footnote{Ref. \onlinecite{oba2008defect}}&0.74 &&\\
LDA\footnote{Ref. \onlinecite{zag14}} &1.24 &1.70&\\
HSE06\footnotemark[2] &2.49&&\\
B3LYP\footnotemark[3] &3.21 &3.49 &\\  \hline
Expt.\footnote{Ref. \onlinecite{Rak13}} & 3.30 &\multicolumn{2}{c}{3.50}\\ \hline
\end{tabular}
\end{center}
\end{table}

We calculated the electronic structure for both polymorphs using DFT-GGA and HSE06, and we also perform $G_0W_0$ and sc$GW$ calculations, using both as starting electronic structure.
Results are shown in Fig.\ \ref{fig:GWband} and the band gaps from all approaches are compared to experimental and other theoretical results in Table\ \ref{tab:band}.
We compare to the HSE06+sc$GW$ band structure, with the gap shifted to the HSE06+$G_0W_0$ value, to illustrate that for ZnO, the sc$GW$ approach merely provides a correction to the band gap, without changing the dispersion of the bands.

As expected, the DFT-GGA band gaps of 0.74 eV (WZ) and 0.9 eV (BN) strongly underestimate experimental results of 3.3 eV (WZ) and 3.5 eV (BN) \cite{Rak13}.
This is well known and was reported in the literature before, however, we note that the \emph{difference} between the band gaps of the two polymorphs (0.16 eV), agrees well between DFT-GGA and experiment.
This confirms earlier findings that relative trends for the electronic structure of similar materials oftentimes are reasonably well reproduced by DFT, even if the absolute value (e.g.\ the band gap) is incorrect.

Band gaps computed using the HSE06 hybrid functional agree much better with experiment for both polymorphs.
Here we use the original HSE06 functional, i.e.\ the mixing parameter is $\alpha$=$0.25$ and the range-separation parameter is  $\omega$=$0.2$ $a_0^{-1}$ \cite{hse,heyd2006erratum}.
While this approach still underestimates the absolute gaps (see Table\ \ref{tab:band}), the difference between both polymorphs is again 0.2 eV.
In addition, our result for the WZ polymorph matches well with a previously reported HSE band gap of 2.5 eV \cite{wrobel09}.

Next, we compute quasiparticle energies by means of the $G_0W_0$ approach.
Using the electronic structure from DFT-GGA as starting point leads to $G_0W_0$ band gaps of 2.10 eV (WZ) and 2.35 eV (BN).
These are about 1.35 eV (WZ) and 1.3 eV (BN) larger than the DFT-GGA band gaps and still underestimate the experimental result (see Table\ \ref{tab:band}).
For a $G_0W_0$ calculation based on the HSE06 electronic structure, the band gaps are 3.22 eV for BN-ZnO and 3.05 eV for WZ-ZnO.
The band-gap differences between both polymorphs are 0.15 eV and 0.17 eV, respectively, similar to what we discussed above.

Next, we compare to the sc$GW$ approach.
In our calculations both the Green's function $G$ and the screened Coulomb interaction $W$ are updated nine times and the difference between the final two steps is smaller than 0.01 eV. 
Band gaps in Table~\ref{tab:band} show that sc$GW$ eliminates the starting-point dependence.
For both polymorphs, the differences of the band gap value between the $G_0W_0$ calculations based on GGA and HSE06 exchange-correlations are about 0.9 eV but when sc$GW$ is performed, those differences are reduced to only about 0.1 eV.
More details about the starting-point dependence and the sc$GW$ calculations are discussed in the supplemental material at [URL will be inserted by publisher]. 
Table~\ref{tab:band} also shows that sc$GW$ calculations lead to larger  gaps compared to experiment, which was reported in the literature before \cite{shishkin2007self}.
This was traced back to the lack of electron-hole interaction in the Coulomb kernel $W$ and the lack of vertex corrections.
A value of 3.9 eV was reported before \cite{van2006quasiparticle} for WZ-ZnO and is in good agreement with our result. 

Finally, we now compare the band gaps computed for BN-ZnO when theoretical and experimental lattice constants are used.
The data in Table \ref{tab:band} clearly shows that using theoretical lattice constants leads to gaps that are about 0.1\,--\,0.3 eV larger than those of WZ-ZnO, across the different computational approaches.
This agrees with another study that reported band gaps for both polymorphs \cite{zag14} and it also agrees with experiment \cite{zag14, Rak13}.
Similarly to what we discussed for WZ-ZnO above, self-consistent $GW$ overestimates the experimental band gap of BN-ZnO by about 10\,\%, when the theoretical lattice constants are used.
Conversely, when using experimental lattice constants, we obtain a band gap more than 1 eV larger than that of WZ-ZnO.
In this case, self-consistent $GW$ overestimates the experimental gap by nearly 50\,\%, which is extremely large.
It is also unusual for sc$GW$ to perform so differently for polymorphs of the same material, by predicting a gap 20\,\% too large for WZ-ZnO, but 50\,\% too large for BN-ZnO.

Instead, we interpret this as an additional indication that the only available experimental lattice constants, reported in Ref.\ \onlinecite{liz10}, are too small.
While the XRD method used in Ref.\ \onlinecite{liz10} is very reliable, the match-up of the XRD pattern with the theoretical pattern of the BN structure is not very clear (see Fig.\ 4 in Ref.\ \onlinecite{liz10}).
The conclusions are essentially based only on three broad peaks in the experimental spectrum.
To fully resolve this problem, more experimental data, new bulk BN-ZnO samples, and improved XRD measurements are needed.

In the following, we will now investigate optical properties of BN-ZnO using our theoretical data for the lattice parameters.
To facilitate this discussion, we introduce the following nomenclature:
The uppermost two valence bands, that are degenerate in both structures at the $\Gamma$ point, are labeled $\Gamma_\textrm{VBM}$.
Crystal-field splitting separates these from the next single non-degenerate valence band, which we denote as $\Gamma_\textrm{CF}$, by 0.09 eV in WZ-ZnO \cite{mang1995band}.
A similar splitting also appears in BN-ZnO, however, the energy difference in this case is much larger (about 0.7 eV) and one set of two-fold degenerate states appears in between (3 meV above $\Gamma_\textrm{CF}$).

\section{\label{sec:opt}Optical properties}

\subsection{\label{sec:spectra}Dielectric functions}

\begin{figure}
\centering
\includegraphics[width=0.48\textwidth]{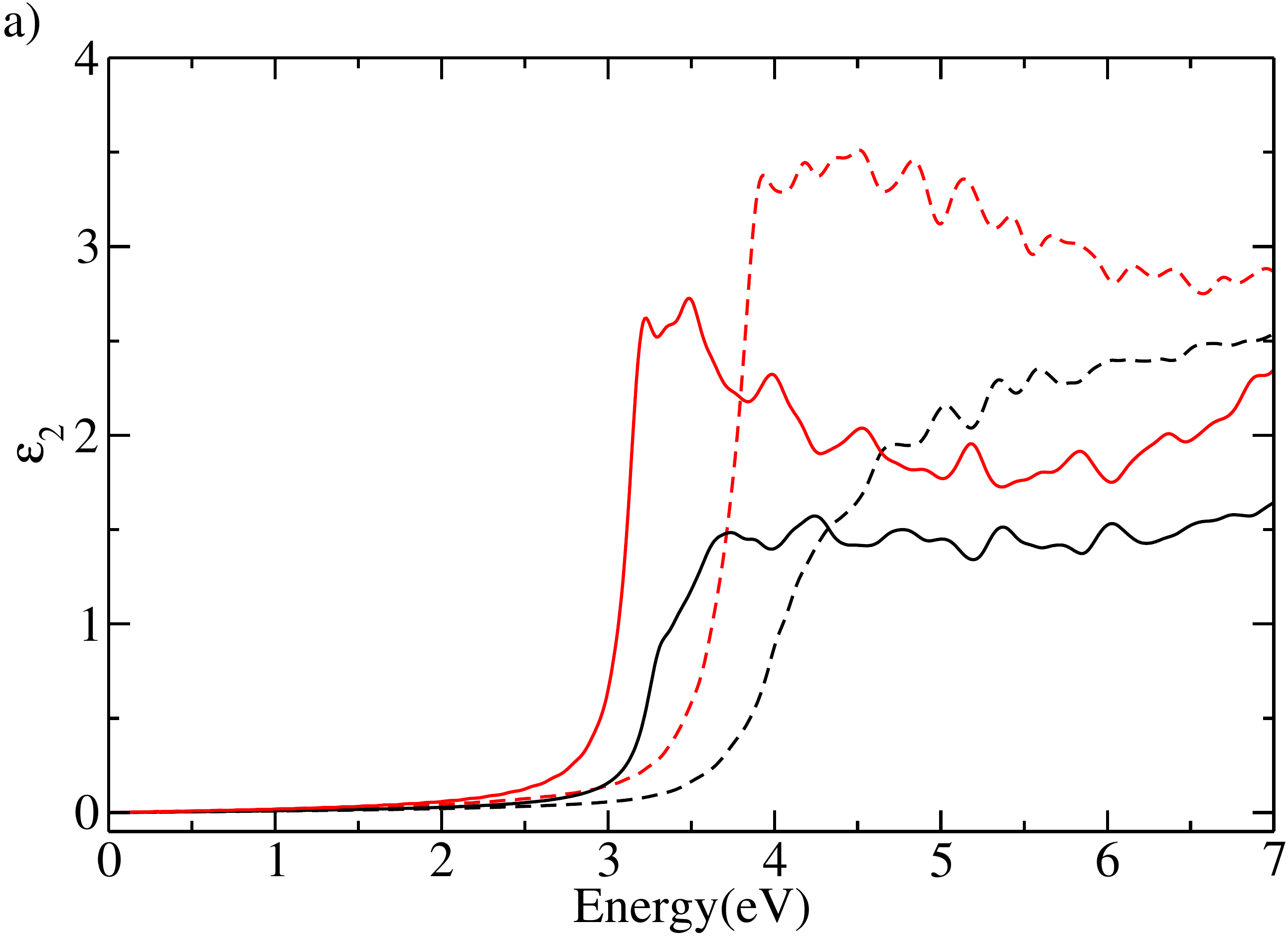}
\includegraphics[width=0.48\textwidth]{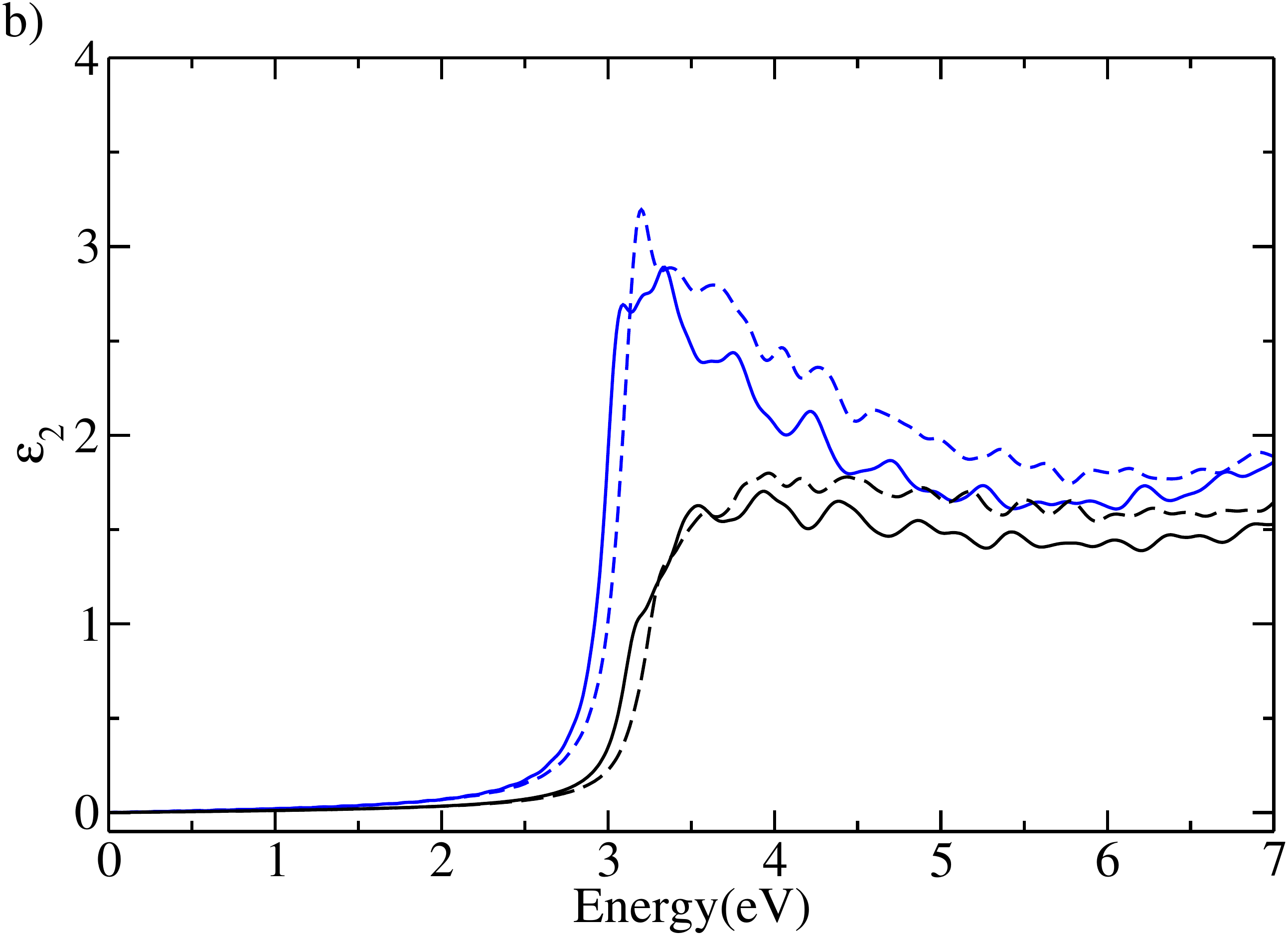}
\caption{\label{fig:dft-bse}(Color online.) Imaginary part of the macroscopic dielectric function, computed using BSE (red, blue) and DFT (black), for (a) BN-ZnO and (b) WZ-ZnO. Solid lines correspond to light polarization perpendicular to the $c$ axis (ordinary) and dashed lines correspond to parallel (extraordinary) polarization.}
\end{figure}

We now report optical properties of WZ-ZnO and BN-ZnO computed using the BSE framework, to investigate the influence of excitonic effects.
The comparison of the optical spectra with (BSE, DFT+scissor) and without (DFT+scissor) excitonic effects shown in Fig.\ \ref{fig:dft-bse} illustrates the influence of electron-hole binding on the spectral shape.
The inclusion of excitonic effects leads to a steeper onset of the spectrum.
In addition, the onset of the BSE spectra occurs at slightly lower (by approximately 0.3 eV) photon energies compared to the DFT+scissor data, which is attributed to exciton binding energies (see discussion in Sec.\ \ref{sec:exci}).
Our result for the optical absorption spectrum of BN-ZnO is a prediction, since there is currently no experimental data available.
In order to illustrate the influence of the crystal structure, we also computed an optical absorption spectrum for WZ-ZnO (see Fig.\ \ref{fig:dft-bse}) and found good agreement with the data in Ref.\ \onlinecite{andre}.

By comparing spectra for both polymorphs
we find a much stronger optical anisotropy for BN-ZnO.
The energy positions of the absorption onsets for ordinary and extraordinary light polarization in BN-ZnO differ by about 0.7 eV, while the difference for WZ-ZnO is not larger than 0.1 eV.
We traced this back to the band structure and optical transition-matrix elements:
From band-resolved optical-absorption spectra (see supplemental material at [URL will be inserted by publisher]) we found that the degenerate $\Gamma_\textrm{VBM}$ states form the absorption onset for ordinary light polarization in both polymorphs (see Fig.\ \ref{fig:dft-bse}).
These $\Gamma_\textrm{VBM}$ states consist of mainly O\,2$p_x$ and O\,2$p_y$ contributions with some Zn\,3$d_{xy}$ and Zn\,3$d_{x^2}$ character.
In contrast, for extraordinary polarization the transition-matrix elements for these states are much smaller, leading to an absorption onset that coincides with transitions from the $\Gamma_\textrm{CF}$ VB into the lowest conduction band.
In WZ-ZnO, this $\Gamma_\textrm{CF}$ state appears 0.1 eV below the VBM and consists mainly of O\,2$p_{z}$ and Zn\,3$d_{z^2}$.
However, as discussed above, in BN-ZnO this $\Gamma_\textrm{CF}$ state appears much lower in energy.
This difference in the energies of the $\Gamma_\textrm{VBM}$ and $\Gamma_\textrm{CF}$ states, along with their characteristic optical matrix elements, leads to the stronger optical anisotropy we find for the BN structure. 

\begin{figure}
\includegraphics[width=0.45\textwidth]{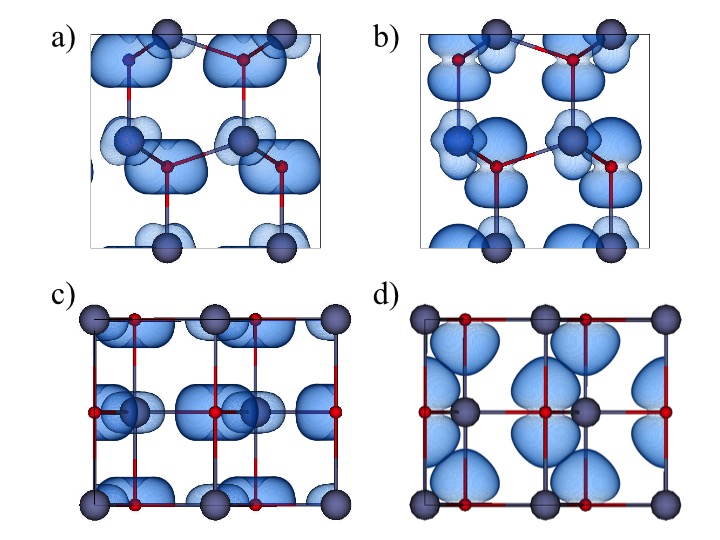}
\caption{\label{fig:wf}[1120] view of Kohn-Sham states at the $\Gamma$ point for (a) $\Gamma_\textrm{VBM}$ in WZ-ZnO, (b) $\Gamma_\textrm{CF}$ in WZ-ZnO, (c) $\Gamma_\textrm{VBM}$ in BN-ZnO, and (d) $\Gamma_\textrm{CF}$ in BN-ZnO. The iso-surface is chosen such that 90\,\% of the electrons lie within.}
\end{figure}

We provide further support to this conclusion by studying the wave function character of these states.
Figures \ref{fig:wf} (a) and (c) show $\Gamma_\mathrm{VBM}$ and Figs.\ \ref{fig:wf} (b) and (d) show $\Gamma_\mathrm{CF}$ for WZ-ZnO and BN-ZnO, respectively.
These plots illustrate the O\,$2p_x$ and O\,$2p_y$ character of $\Gamma_\textrm{VBM}$ and the O\,$2p_z$ character of $\Gamma_\textrm{CF}$ for both materials, which leads to the different polarization dependence.

\begin{figure}
\includegraphics[width=0.45\textwidth]{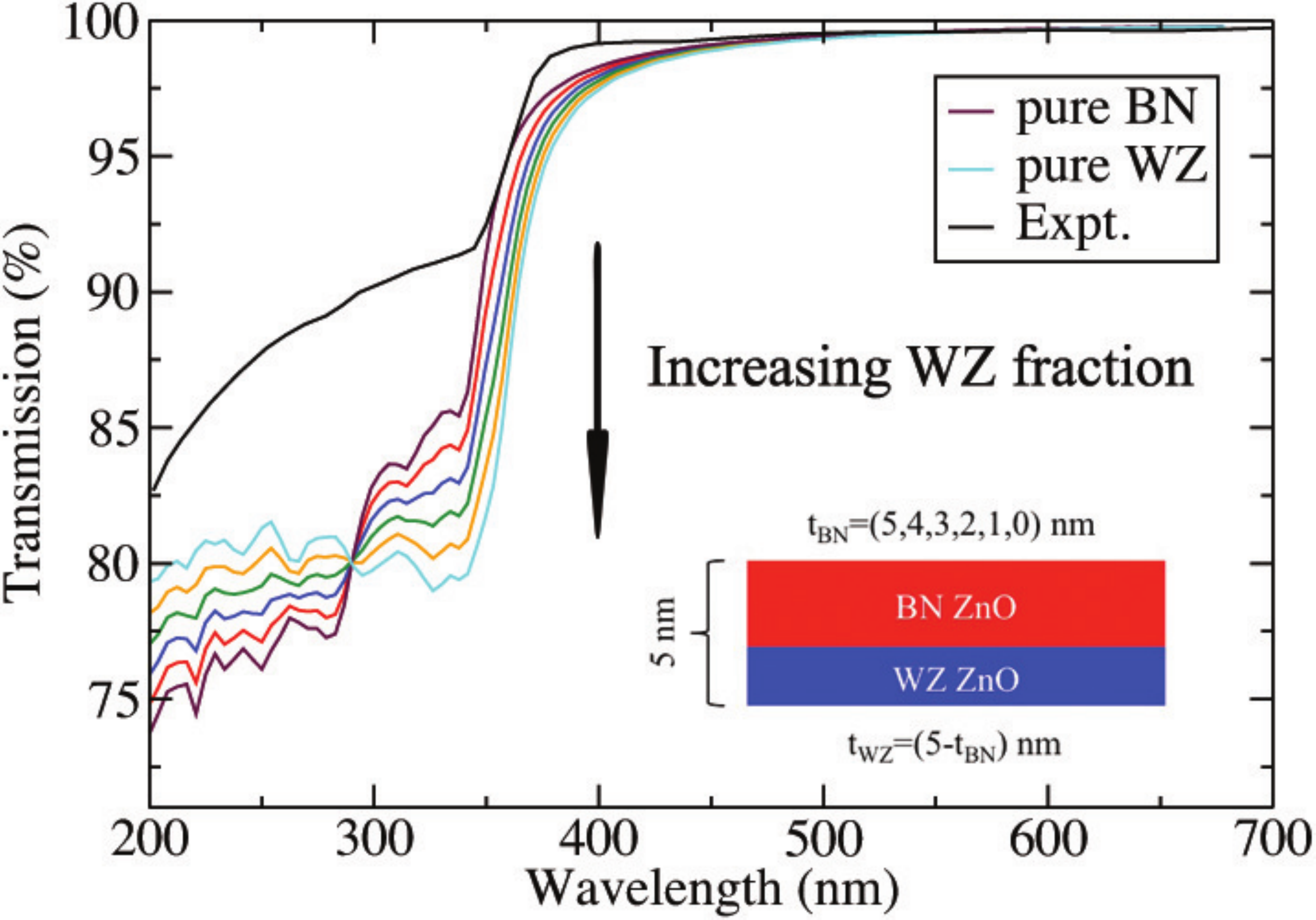}
\caption{\label{fig:maxwell}Maxwell simulation results for a 5 nm thin film with different fractions of BN-ZnO and WZ-ZnO. The arrow indicates increasing WZ-ZnO fraction in steps of 1 nm between two successive curves. The inset of the figure shows the structure simulated schematically.}
\end{figure}

Finally, we explore the influence of BN-ZnO on the optical properties of experimentally realized ultra-thin ZnO films \cite{chi2017ultra}. 
To this end, we applied the transfer-matrix method \cite{heavens1960optical, orfanidis2002electromagnetic} in Matlab \cite{matlab14} to compute optical properties of a 5 nm thin film that contains different fractions of BN-ZnO and WZ-ZnO.
This is on the order of magnitude at which BN-ZnO can appear \cite{tusche07,claeyssens2005growth, wu2011stabilizing}.
From these simulations we compute the transmittance and in Fig.\ \ref{fig:maxwell} we compare with experimental data from the work of Chi \emph{et al.}\cite{chi2017ultra}
Our results are blue-shifted by 0.45 eV to match the position of the absorption onset with the experimental data for the thin films.

This figure illustrates the distinct optical signatures of both polymorphs in the spectral region around the optical absorption onset:
Due to the strong anisotropy, BN-ZnO shows a two-step onset at photon wave lengths of 280 nm and 340 nm. 
WZ-ZnO, on the other hand, only has one onset and a valley for transmission around wave lengths of 340 nm.
The experimental data shows increasing transmission across wave lengths from 200 nm to 350 nm. 
Comparison to our data shows that the behavior observed in experiment is much more consistent with BN-ZnO, or with mixtures of both polymorphs, instead of pure WZ-ZnO (see Fig.\ \ref{fig:maxwell}).
We note that in our simulations of the thin film, we assumed a sharp interface between BN-ZnO and WZ-ZnO, neglecting possible atomic rearrangements.
Despite of this, since multiple theoretical studies \cite{kulkarni06novel,Rak13,wu2011stabilizing} predicted a  transition from WZ-ZnO to BN-ZnO in thin films under tensile strain, and since our simulations clearly show distinctly different optical-transmission signatures around the absorption onset, we raise the possibility that BN-ZnO may have affected thin films in experiment without being explicitly noticed \cite{chi2017ultra}, and we suggest further experiments to clarify this.
In addition, our results show that purposefully tuning the different fractions of BN-ZnO and WZ-ZnO may provide a feasible route towards tunable optical properties in the UV region of ZnO thin films and nano-structures.

\subsection{\label{sec:exci}Exciton-binding energies}

Due to the Coulomb interaction between electrons and holes, the lowest optical excitation energy is smaller than the single-quasiparticle band gap.
The difference between these two quantities is the exciton-binding energy \cite{sham1966many,hanke1979many} and is critically important to understand, for instance, the strength of optical absorption of a material or the splitting of electron-hole pairs after absorption of light.
In many crystalline semiconductors, dielectric screening of the Coulomb interaction causes the binding energy to be on the order of 10\,--\,100 meV \cite{wannier1937structure,kittel1976introduction,elliott1957intensity}.

In ZnO, the first conduction band is very dispersive and has parabolic character in the close vicinity of the $\Gamma$ point, which leads to Wannier-Mott type excitons that can be studied analytically using a parabolic two-band model.
Within this model, the exciton-binding energy is determined by the excitonic Rydberg \cite{fuchs2008efficient}, 
\begin{equation}
\label{eq:ewm}
R_\mathrm{ex}=R_{\infty}\mu/(m_0\varepsilon_0^2),
\end{equation}
where $R_{\infty}$ is the Rydberg constant ($R_{\infty}$=$-13.6$ eV), $\varepsilon_0$ is the electronic dielectric constant, $\mu$ is the reduced electron-hole mass ($\mu^{-1}$=$m^{*-1}_e$+$m^{*-1}_h)$, and $m_0$ is the free-electron mass.

We obtained the electronic dielectric constant using DFT-GGA and the independent-particle approximation.
This yields a value of $\varepsilon_0$=$5.11$ for BN-ZnO and $\varepsilon_0$=$5.25$ for WZ-ZnO (see Table \ref{tab:dielectric}).
The effective electron and hole masses were computed via parabolic fits \cite{simon2013oxford} to the GGA band structure, which is much better resolved in $\mathbf{k}$-space than HSE06 or sc$GW$ results.
We estimate the error introduced by using GGA over sc$GW$ data for the effective-mass fit to be on the order of 7\,\%. 

\begin{table}
\begin{center}
\caption{\label{tab:efm}Effective mass fits (to DFT-GGA) and exciton-binding energies computed within the Wannier-Mott model for different $\mathbf{k}$ ranges along the $\Gamma \rightarrow M$ direction. The first column corresponds to the fraction of the fitted range in reciprocal space of the full $\Gamma \rightarrow M$ length.}
\centering{Effective mass fit for BN structure\label{wzefmas}}
{\begin{tabular}{>{\centering\arraybackslash}p{1in}cccc}
\hline
$\mathbf{k}$-range of fitting & $m^*_e/m_0$  & $m^*_h/m_0$ &$\mu/m_0$ &$R_\mathrm{ex}$ (meV) \\ \hline
0.0375	&0.153&    1.671	&  0.140 & 73.0 \\
0.0500	&0.164	&	1.665	&0.149&77.7\\
0.0625     &0.176	     &	1.661 &0.159&83.1\\
0.0750	&0.190	&1.659&0.171&88.9\\
0.0875	&0.204	&1.656&0.182&94.7\\ \hline
\end{tabular}}\\
\centering{Effective mass fit for WZ structure\label{bnefmas}}
{\begin{tabular}{>{\centering\arraybackslash}p{1in}cccc}
\hline
$\mathbf{k}$-range of fitting & $m^*_e/m_0$  & $m^*_h/m_0$ &$\mu/m_0$ &$R_\mathrm{ex}$ (meV) \\ \hline
0.0375	&0.150&    2.703	&  0.143 & 70.3 \\
0.0500	&0.164	&	2.707	&0.155 &76.6\\
0.0625       &0.181	     &	2.719 &0.169&83.6\\
0.0750	&0.198	&2.722&0.185&91.1\\
0.0875	&0.216	&2.735&0.200&98.8\\ \hline
\end{tabular}}
\end{center}
\end{table}

One remaining source of uncertainty when using the Wannier-Mott picture is the non-parabolic character of the lowest conduction band of ZnO.
This introduces a strong dependence of the effective mass on the $\mathbf{k}$-point range used for the fit as points further away from $\Gamma$ are included.
We quantify the effect of this on the exciton binding energy in Table~\ref{tab:efm}.
Our results show that changing the fitting range from 3.75\,\% of the full BZ length along the $\Gamma \rightarrow M$ direction to 8.75\,\% causes an increase in the electron effective mass by 33\,\% (44\,\%) for BN-ZnO (WZ-ZnO). 
Since the exciton binding energy is directly proportional to the reduced mass, it is also highly sensitive to this change (see Table~\ref{tab:efm}). 
Using a linear extrapolation, we obtained exciton-binding energies from the Wannier-Mott model of 56.2 meV for BN-ZnO and 48.3 meV for WZ-ZnO. 
In this analysis we used only the effective mass along the $\Gamma \rightarrow M$ direction.
In the following, we overcome this limitation of the Wannier-Mott picture by solving the BSE to compute exciton-binding energies from first principles, using the full band structure of the material instead of parabolic fits.

To accurately compute exciton-binding energies on the order of several ten meV, a high $\mathbf{k}$-point density is needed for sampling the Brillouin zone (BZ) \cite{kittel1966introduction, fuchs2008efficient} and a large BSE energy cutoff, i.e., the highest-energy electron-hole excitation included in the exciton Hamiltonian.
We found that exciton binding energies are converged to within 2\,\% for a BSE energy cutoff of 6 eV (see supplemental material at [URL will be inserted by publisher]).
To converge BZ sampling at acceptable computational cost, we use hybrid $\mathbf{k}$-point grids up to $8\,\times\,8\,\times\,6:3\,\times\,3\,\times\,3:85.3\,\times\,85.3\,\times\,44$ in this work, following the notation in Ref.\ \onlinecite{fuchs2008efficient}.
Linear extrapolation (details see supplemental material at [URL will be inserted by publisher]) then yields converged exciton binding energies.
We compute 63.8 meV for BN-ZnO and 51.3 meV for WZ-ZnO without correcting for remaining BSE energy cutoff convergence, which would increase these numbers by about 4 meV.

These values are larger than those obtained using the Wannier-Mott model.
This is because in a real material more than two bands contribute to exciton formation and those bands are neither strictly parabolic nor isotropic.
Hence, the binding energies computed from converged first-principles BSE calculations are more reliable. 
Experiments report the exciton-binding energy for WZ-ZnO to be about 60 meV \cite{janotti2009fundamentals}.
This is slightly larger than but close to our value.

\begin{table}
\caption{\label{tab:dielectric}Electronic dielectric constants for WZ-ZnO and BN-ZnO, both for ordinary and extraordinary light polarization.}
\begin{tabular}{cccc}
 &$\varepsilon_{\perp c}$&$\varepsilon_{\parallel c}$&$\varepsilon_\textrm{avg}$\\ \hline
BN-ZnO &5.12&5.09&5.11\\
WZ-ZnO &5.22&5.29&5.25\\
Comp. (WZ-ZnO)\cite{schleife2006first} & 5.24 & 5.26 &5.25\\
Expt. (WZ-ZnO)\cite{ashkenov2003infrared} & 3.70 &3.78 & 3.73\\
\hline
\end{tabular}
\end{table}

When comparing exciton-binding energies for both polymorphs, it is apparent that the value for BN-ZnO is about 10 meV larger, which indicates that excitonic effects are slightly stronger in BN-ZnO.
In order to explain this behavior we investigate the parameters that enter the Wannier-Mott model:
The data in Table~\ref{tab:efm} shows that the reduced electron-hole masses differ by about 2\,\% (0.140 for BN-ZnO and 0.143 for WZ-ZnO) when fitted only in the closest vicinity of the BZ center.
The differences are also about 2\,\% for the dielectric constants of both polymorphs (see Table~\ref{tab:dielectric}), however, as can be seen from Eq.\ \eqref{eq:ewm}, this parameter enters quadratically into the exciton-binding energy.
Thus, this difference in dielectric screening is the main reason for the larger excitonic effects observed for BN-ZnO.
Finally, when comparing the dielectric constant for WZ-ZnO to earlier computational results \cite{schleife2006first}, we find good agreement.
The computational value is smaller than the experimental result, which is due to the underestimation of the band gap in DFT-GGA.

\section{\label{sec:conclusion}Conclusions}

In this work we used cutting-edge first-principles simulations, based on many-body perturbation theory, to provide a consistent picture of structural, electronic, and optical properties of the non-equilibrium BN-ZnO polymorph.
Our results for lattice constants of WZ-ZnO and BN-ZnO are in good agreement with previous computational work, and we find good agreement with experiment for the WZ phase.
Lattice constants calculated for BN-ZnO overestimate experiment by about 10\,\%, as reported before.
While this is an unusually large discrepancy, we show that the BN-ZnO band gap is about 0.2 eV larger than the WZ-ZnO band gap when these atomic coordinates are used.
This is in excellent agreement with experiment and we find that this result does not depend on the description of exchange and correlation.

We further report optical absorption spectra, including excitonic effects, for both polymorphs.
While the results for WZ-ZnO agree very well with previous data, we predict a large optical anisotropy for the BN structure.
Exciton binding energies are 63.8 meV (BN-ZnO) and 51.3 (WZ-ZnO).
Further analysis using a parabolic two-band model shows that the exciton binding energy for BN-ZnO is about 10 meV larger due to the smaller dielectric constant of this polymorph, which we associate with the difference in bond lengths.
Finally, we employ transfer-matrix method Maxwell simulations to study the influence of BN-ZnO contributions on the optical properties of ultra-thin films.
We find that the optical-transmission signature in the ultraviolet spectral region is characteristic for each crystal structure.

Our results provide the first \emph{consistent} picture of structural, electronic, and optical properties of the BN polymorph of ZnO.
The band-gap difference between BN-ZnO and WZ-ZnO is in agreement with experiment, close to 0.2 eV, when the theoretical lattice geometry is used.
In contrast, this difference amounts to more than 1 eV when the experimental lattice constants are used.
Hence, we conclude that, in the absence of more accurate experimental data for the lattice geometry, the theoretical lattice geometry is most accurate and resolves this long-standing controversy.
At the same time, this observation makes a strong case for further experimental work to accurately explore the lattice geometry of bulk BN-ZnO.

More broadly, our work is an example for first-principles simulations that provide detailed information on optical properties, accurate enough to allow direct conclusions about structural properties and even phase identification. 
We clearly identify optical signatures of both polymorphs and by comparing with experimental data for 5 nm thin films we highlight the possibility that BN-ZnO is present in previously studied nanostructures, without being explicitly noticed.
Again, we hope that our results lead to more experimental efforts in this direction.

Finally, we explicitly highlight that our data has strong experimental implications, e.g.\ when the fraction of BN-ZnO and WZ-ZnO in a given sample can be purposefully designed.
This directly allows for tuning of optical properties, which is important for nanostructures and thin films, e.g.\ for UV detectors and UV protection.

\begin{acknowledgments}
We thank Axel Hoffmann and Emmanouil Kioupakis for fruitful discussions.
This material is based upon work supported by the National Science Foundation under Grant No.\ DMR-1555153.
This research is part of the Blue Waters sustained-petascale computing project, which is supported by the National Science Foundation (awards OCI-0725070 and ACI-1238993) and the state of Illinois. Blue Waters is a joint effort of the University of Illinois at Urbana-Champaign and its National Center for Supercomputing Applications.
\end{acknowledgments}

\bibliography{./ref}

\newpage

\section*{Supplemental Material}

\subsection{\label{sup:ev} Convergence test for the structural relaxation}

\begin{figure}[h]
\begin{center}
\includegraphics[width=0.5\textwidth]{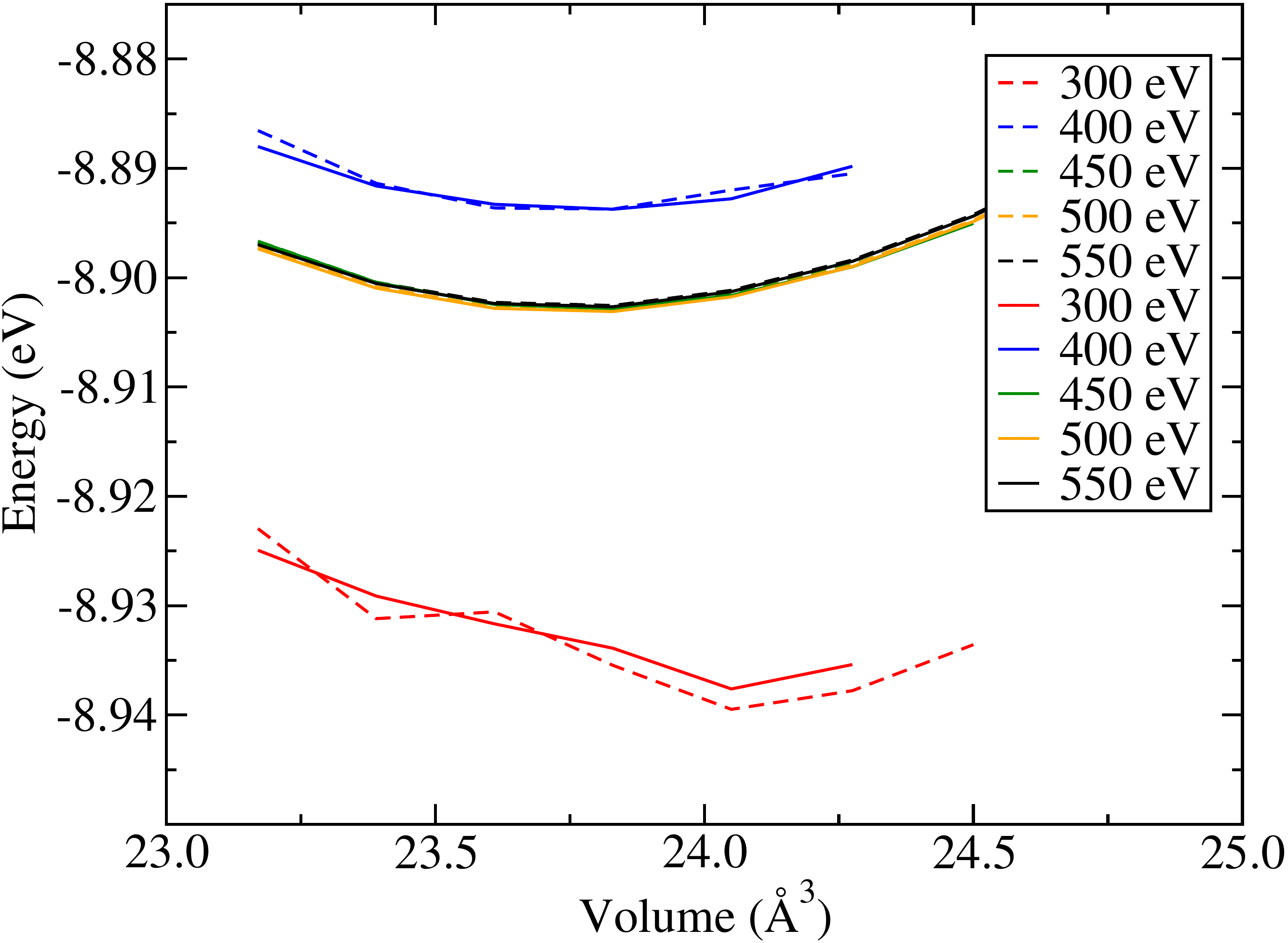}
\caption{\label{fig:bnev}(Color online.) Total energy as a function of the volume (both per formula unit) for the relaxation of the BN structure of ZnO. In the figure, dashed lines represent results for $8\,\times\,8\,\times\,6$ $\mathbf{k}$ points and solid lines represent $12\,\times\,12\,\times\,8$. Different colors represent different plane-wave cutoffs.}
\end{center}
\end{figure}

The energy vs.\ volume curve for structural relaxation is shown in Fig.\ \ref{fig:bnev} and the corresponding data is given in Table~\ref{chart:relax}.
We have tested for different plane-wave cutoff energies and $\mathbf{k}$-point samplings to make sure that the total energy is well converged.
It can be seen from the figure that changing the plane-wave cutoff could have a very large influence on the total energy of the unit cell.
Starting from 450 eV, the total energy begins to converge.
We tested $8\,\times\,8\,\times\,6$ and $12\,\times\,12\,\times\,8$ $\mathbf{k}$ points.
Using $12\,\times\,12\,\times\,8$ yields  a smoother curve with less noise.
Thus we have chosen a plane-wave cutoff of 450 eV and a $12\,\times\,12\,\times\,8$ $\mathbf{k}$-point grid for our calculations. 

\begin{table}[h]
\caption{\label{chart:relax} The total energy for different convergence parameters and different volume  for BN ZnO. The blanks comes from that for some of the volumes, the BN cell will relax to the WZ so that we cannot get BN structure in these volumes. The E-V curve we get will be approximately parabolic if we are close to the minimum and have a well-converged k-points mesh and energy cutoff. }
\begin{center}
\begin{tabular}{cccccccccccccc}
\hline
K points      & \multicolumn{5}{c}{$8\,\times\,8\,\times\,6$}                                      \\ \hline
Cutoff (eV) & 300        & 400        & 450        & 500        & 550        \\ \hline
Volume (\r{A})        & \multicolumn{5}{c}{Total Energy (eV)}                         \\ \hline
46.34         & $-$17.845 & $-$17.773 & $-$17.793 & $-$17.794  & $-$17.793 \\
46.78         & $-$17.862 & $-$17.782 & $-$17.800 & $-$17.802 & $-$17.801  \\
47.22         & $-$17.861 & $-$17.787 & $-$17.805 & $-$17.805 & $-$17.805 \\
47.66         & $-$17.871 & $-$17.787 & $-$17.806 & $-$17.806 & $-$17.805 \\
48.10         & $-$17.879 & $-$17.784 & $-$17.803 & $-$17.803 & $-$17.802 \\
48.55         & $-$17.876 & $-$17.781 & $-$17.798 & $-$17.798 & $-$17.797 \\
49.00         & $-$17.867 &                    & $-$17.790 & $-$17.790   & $-$17.788  \\
49.45         &            	     &                    &                    & $-$17.779 & $-$17.778 \\ \hline
\end{tabular}
\begin{tabular}{cccccccccccccc}
\hline
K points      & \multicolumn{5}{c}{$12\,\times\,12\,\times\,8$}                                    \\ \hline
Cutoff (eV) & 300        & 400        & 450        & 500        & 550        \\ \hline
Volume (\r{A})        & \multicolumn{5}{c}{Total Energy (eV)}                         \\ \hline
46.34         & $-$17.850 & $-$17.776 & $-$17.793 & $-$17.795 & $-$17.794 \\
46.78         & $-$17.858 & $-$17.783  & $-$17.801 & $-$17.802 & $-$17.801 \\
47.22         & $-$17.863 & $-$17.787 & $-$17.805 & $-$17.806 & $-$17.805 \\
47.66         & $-$17.868 & $-$17.787 & $-$17.806 & $-$17.806 & $-$17.805 \\
48.10         & $-$17.875 & $-$17.786 & $-$17.803 & $-$17.804  & $-$17.803  \\
48.55         & $-$17.871 & $-$17.780 & $-$17.798 & $-$17.798 & $-$17.797 \\
49.00         &            &            & $-$17.790 & $-$17.790 & $-$17.789 \\
49.45         &            &            &            & $-$17.779 & $-$17.778  \\ \hline
\end{tabular}
\end{center}
\end{table}

\subsection{\label{sup:scgw} Dependence of self-consistent $GW$ calculations on the starting point}

\begin{figure}[h]
\includegraphics[width=0.45\textwidth]{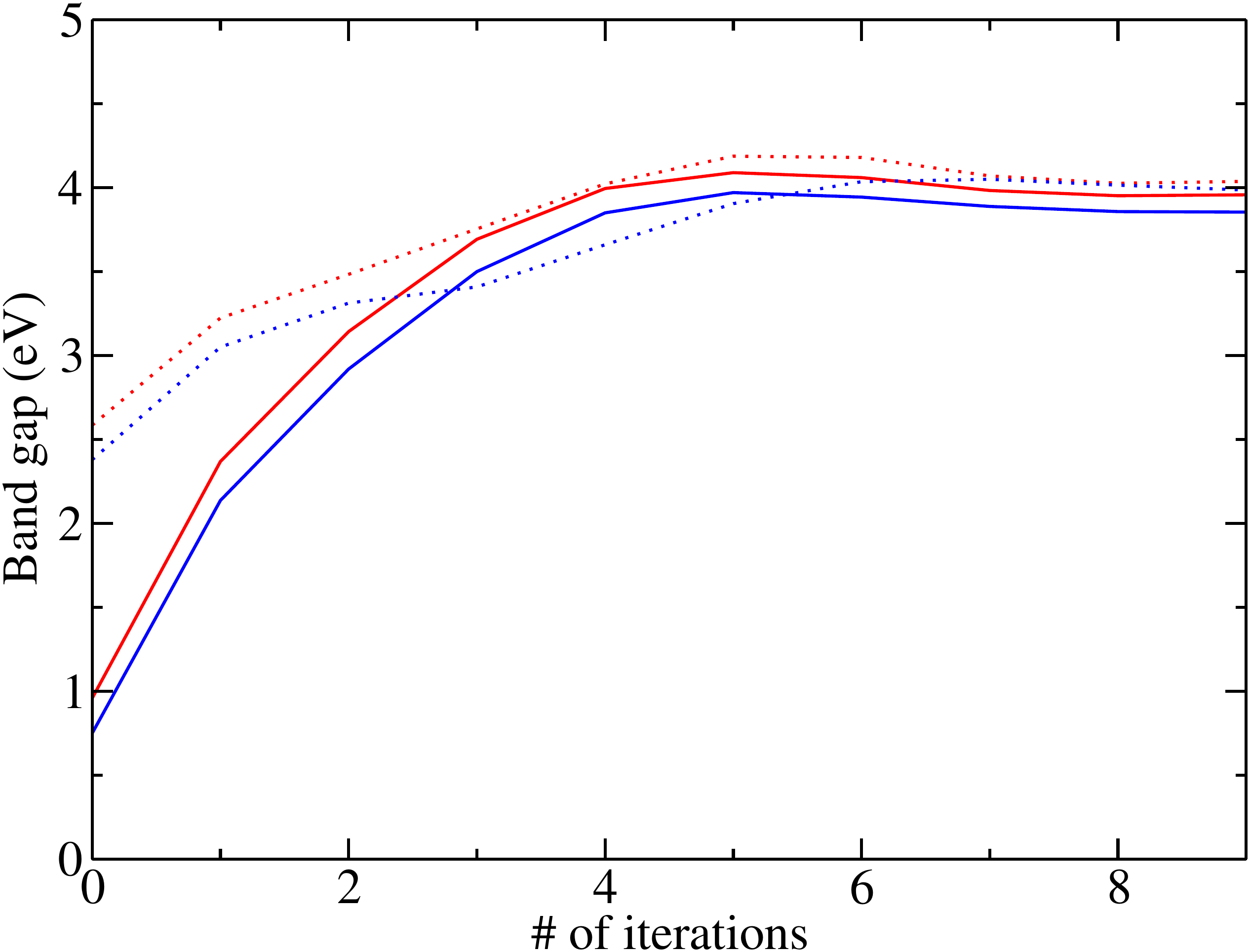}
\caption{\label{fig:stpt}The test for the dependence of the sc$GW$ band gap on the starting wave function. Red represents the results for BN-ZnO and blue WZ-ZnO. Solid lines represent calculations based on GGA and dashed lines represent calculations based on HSE06 input.}
\end{figure}

We have tested for the dependence of self-consistent $GW$ (sc$GW$) calculations on the starting point.
We calculated the sc$GW$ band gap based on DFT wave functions using both GGA exchange correlation and HSE06 exchange correlation for both BN structure and WZ structure.
In our calculations, we performed iteration of both $G$ and $W$ for a total of nine steps.
The results are shown in Fig.\ \ref{fig:stpt}. 
The $G_0W_0$ band gap is 2.37 eV (based on GGA) and 3.22 eV (based on HSE06) for BN-ZnO.
The difference is 0.85 eV.
For WZ-ZnO this difference is even larger at 0.91 eV.
However, with the inclusion of nine steps of self consistency, the difference between $GW$ gap based on GGA wave functions and HSE06 wave functions is reduced to 0.08 eV for BN-ZnO and 0.13 eV for WZ-ZnO.
These numbers are only 2\,\% and 3.2\,\% of the calculated band gaps.
This result verifies our argument that sc$GW$ will no longer dependent strongly on the starting point.

\subsection{\label{sup:BSEenergy} Convergence tests for BSE calculations}

\begin{figure}[h]
\begin{center}
\includegraphics[width=0.48\textwidth]{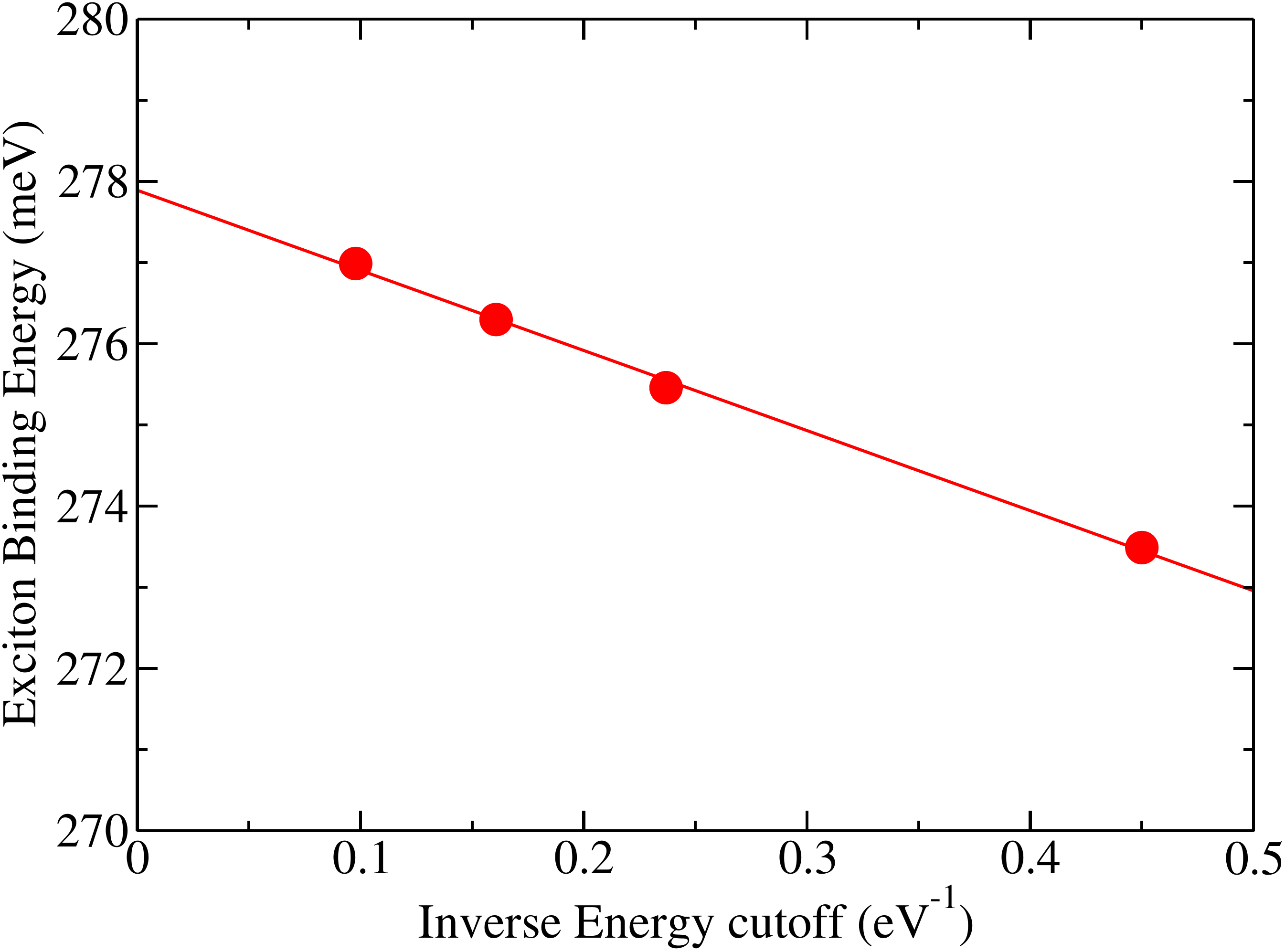}
\caption{\label{fig:energy}Convergence of the exciton binding energy with respect to the inverse BSE energy cutoff. The $\mathbf{k}$-point sampling used is $8\,\times\,8\,\times\,6$.}
\end{center}
\end{figure}

When performing BSE calculations, in principle we should consider transitions between all valence and conduction bands.
However, due to the limit of computational cost, in practice we only consider transitions within a certain energy range, which is specified as the BSE cutoff energy.
This cutoff describes the largest single-particle excitation energy of any electron-hole pair included in computing the exciton Hamiltonian.
We perform BSE calculations using a $\mathbf{k}$-point grid of $8\,\times\,8\,\times\,6$ for BN-ZnO and vary the BSE energy cutoff.
The result is shown in Fig.\ \ref{fig:energy}.
This shows that the calculated exciton-binding energy decreases as the energy cutoff of the BSE calculation increases (as the inverse decreases).
As reported in Ref.\ \onlinecite{fuchs2008efficient}, it shows a nearly linear behavior vs.\ $\{E_\mathrm{cutoff}$-$E_\mathrm{g}\}^{-1}$.
Our tests also show that the error for the exciton-binding energy is within 2\,\% of the converged value (the value linearly extrapolated to 0.0 eV$^{-1}$ in Fig.\ \ref{fig:energy}) for a BSE cutoff of 6 eV.

\begin{figure}[h]
\begin{center}
\includegraphics[width=0.48\textwidth]{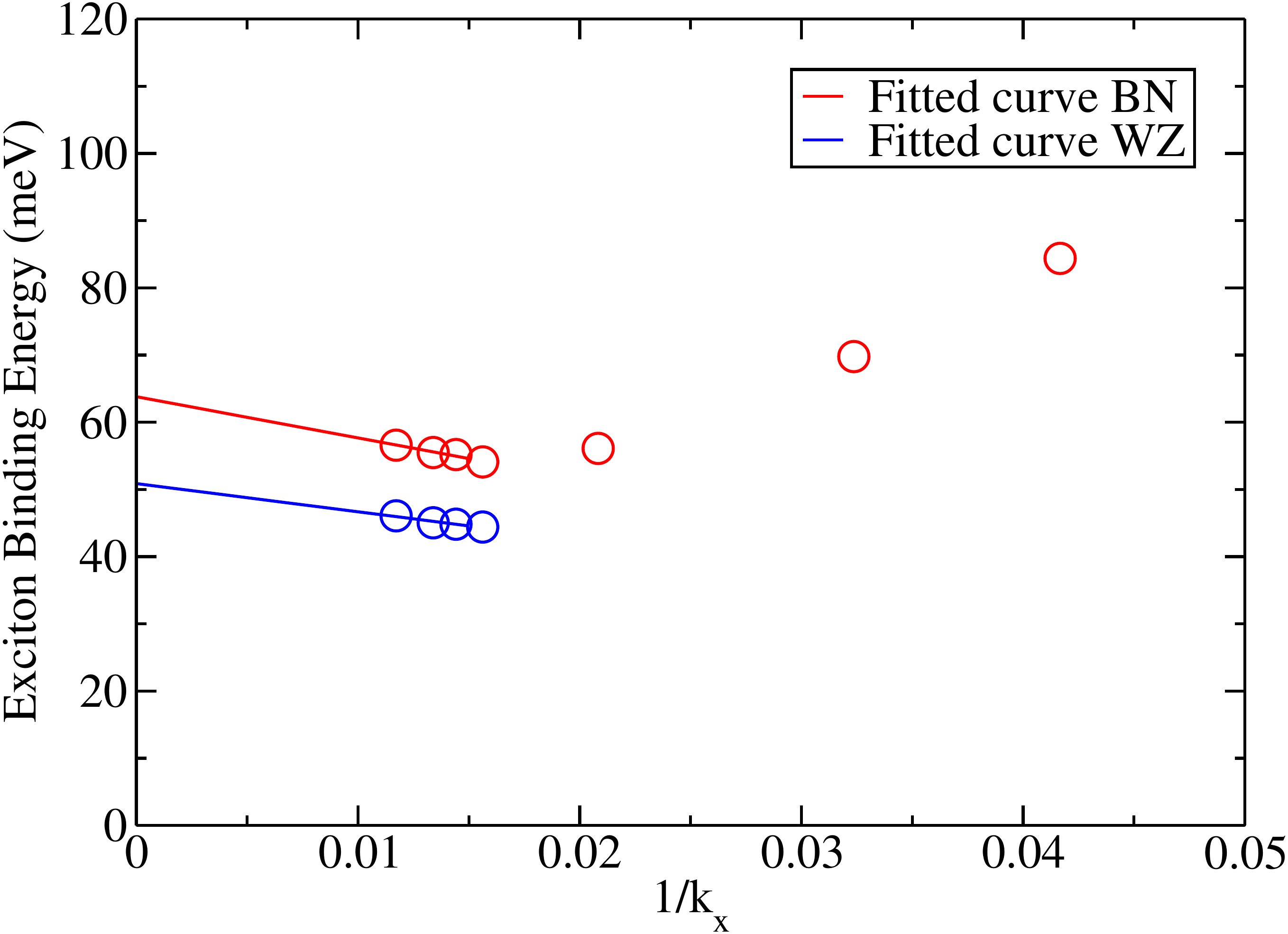}
\caption{\label{fig:kpcon}Convergence of the exciton binding energy with respect to $\mathbf{k}$-point sampling. The BSE energy cutoff is fixed at 6 eV. The graph shows that the ``turn-around'' point for ZnO is at around 1/k$_x$=0.015, which is still an extremely dense sampling, even with the hybrid-mesh scheme. The extrapolated value of the exciton binding energy is 63.8 meV for BN-ZnO and 51.3 meV for WZ-ZnO.}
\end{center}
\end{figure}

For small enough minimum $\mathbf{k}$-point distances the exciton-binding energy starts to depend linearly on the $\mathbf{k}$-point distance.
Hence, convergence tests are needed.
To reduce the computational cost as much as possible, we used hybrid $\mathbf{k}$-point grids \cite{fuchs2008efficient}.
The hybrid mesh samples the BZ non-uniformly, with a higher density near the $\Gamma$ point in the center and lower density away from the center.
Since the lowest eigenenergy for both ZnO polymorphs appears at the $\Gamma$ point, this method provides a way to effectively increase the sampling density near the region of interest, with smaller total numbers of sampling points needed, compared to uniformly sampling the entire BZ.

The plot for converging the exciton binding energies with respect to $\mathbf{k}$-point sampling is shown in Fig.\ \ref{fig:kpcon}.
Our observed behavior matches that reported in Ref.\ \onlinecite{fuchs2008efficient}:
As the number of $\mathbf{k}$ points in a certain direction increases, we find a decrease of the binding energy, followed by a ``turn-around" point.
For even higher samplings, we see a linear region (see Fig.\ \ref{fig:kpcon}).
For BN-ZnO, very large sampling densities are required to observe the turn-around point and to reach the linear region.
For WZ-ZnO we compute the binding energy only for the four best converged samplings. 

\subsection{\label{sup:opt} Band resolved optical spectra}

\begin{figure}[h]
\includegraphics[width=0.45\textwidth]{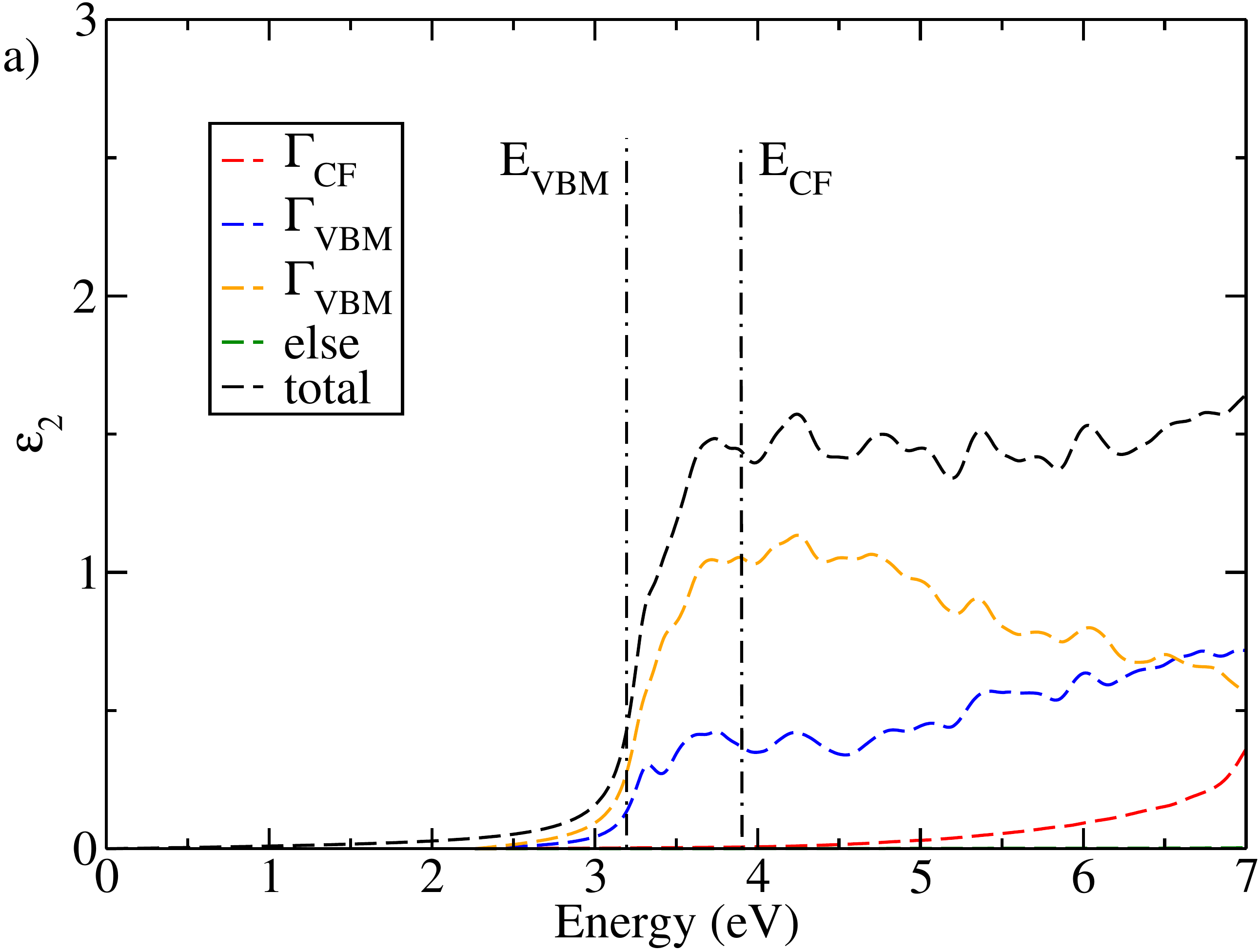}\\
\includegraphics[width=0.45\textwidth]{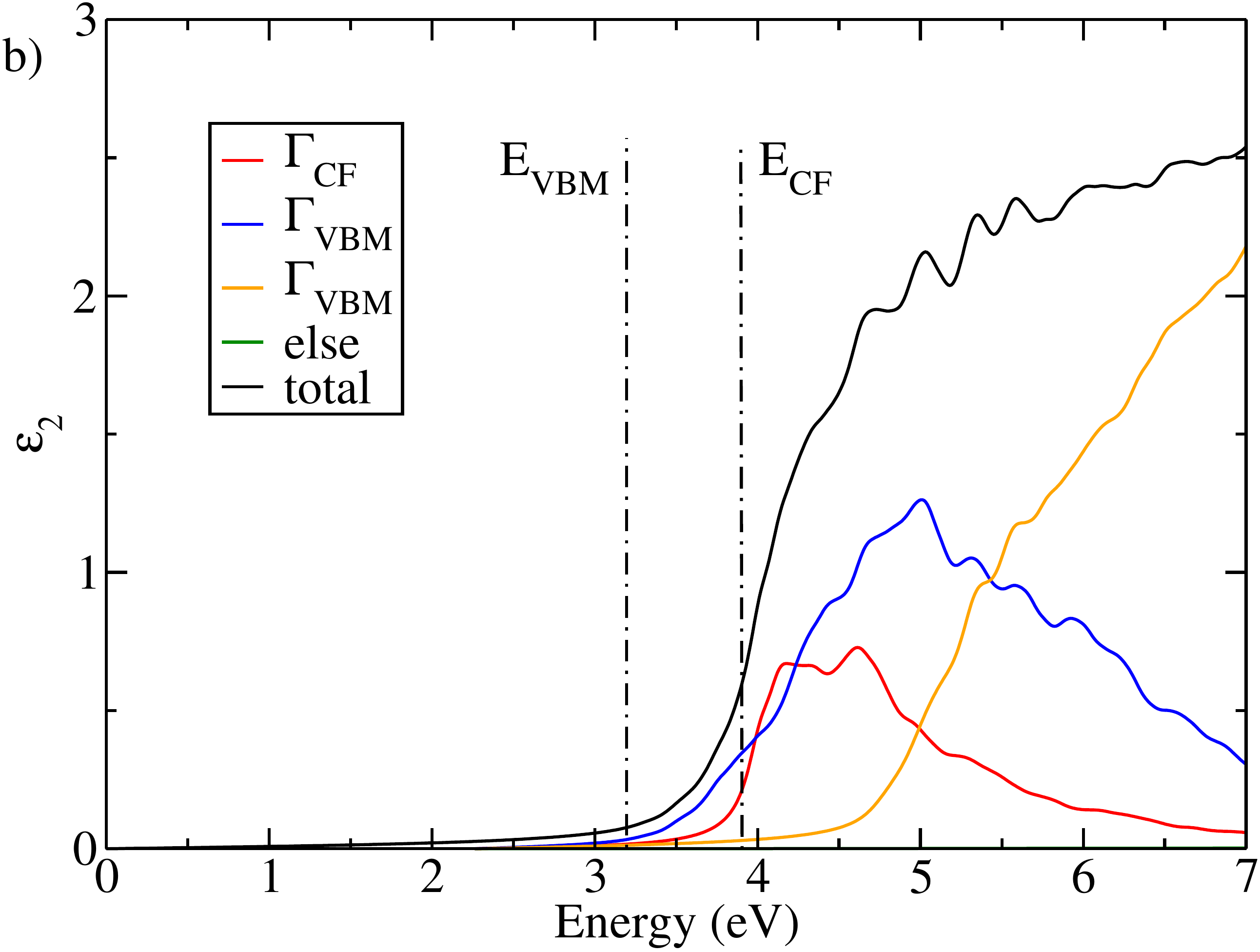}\\
\caption{\label{fig:bandresolved}Band-resolved dielectric function for BN-ZnO. Dashed lines are the band-resolved spectra for ordinary light polarization and solid lines are the spectra for extraordinary light polarization. Dot-dashed lines indicate the energies of transitions from $\Gamma_\textrm{VBM}$ and $\Gamma_\textrm{CF}$ into the conduction-band minimum, respectively.}
\end{figure}

In order to understand what is causing the optical anisotropy of BN-ZnO, we have plotted the optical spectra, resolved with band indices. 
The result is shown in Fig.\ \ref{fig:bandresolved}.
We can see that for BN-ZnO there are two major contribution to the anisotropy:
The first contribution is the optical transition-matrix element.
The valence band maximum is a two-fold degenerate state at $\Gamma$ and we label these as $\Gamma_\textrm{VBM}$ also when they are non-degenerate away from $\Gamma$.
The transition-energy onset of these two states appears around 3.2 eV, but the spectrum increases very slowly for extraordinary polarization.
For ordinary polarization it starts right at the transition energy.
This is due to small optical transition-matrix elements for extraordinary polarization.
The second contribution is the band splitting.
The $\Gamma_\textrm{CF}$ band appears only 0.09 eV below the VBM for WZ-ZnO.
However, in BN-ZnO, this band appears 0.7 eV below the VBM.
This is the band where the optical transition-matrix elements start to become larger for extraordinary polarization.

\end{document}